\documentclass[11pt,a4paper]{article}
\usepackage{jheppub_kim}

\usepackage{pdflscape}
\usepackage{amsmath}
\usepackage{amssymb}
\usepackage{dcolumn}
\usepackage{bm}
\usepackage{color}
\usepackage{epsfig}
\usepackage{amsfonts}
\usepackage{graphicx}
\usepackage{subfigure}
\usepackage{dcolumn}

\newcommand{\be}{\begin{equation}}
\newcommand{\ee}{\end{equation}}
\newcommand{\bea}{\begin{eqnarray}}
\newcommand{\eea}{\end{eqnarray}}

\def\be{\begin{equation}}
\def\ee{\end{equation}}
\def\bea{\begin{eqnarray}}
\def\eea{\end{eqnarray}}

\begin{document}

\title{Holographic superconductor in a deformed four-dimensional STU model}
\author[a]{B. Pourhassan}
\author[a]{M.M. Bagheri-Mohagheghi }

\affiliation[a]{School of Physics, Damghan University, Damghan, 3671641167, Iran}

\emailAdd{b.pourhassan@du.ac.ir}
\emailAdd{bmohagheghi@du.ac.ir}

\abstract{In this paper, we consider deformed STU model in four dimensions including both electric and magnetic charges. Using AdS/CFT correspondence, we study holographic superconductor and obtain transport properties like electrical and thermal conductivities. We obtain transport properties in terms of the black hole magnetic charge and interpret it as magnetic monopole of dual field theory. We find that presence of magnetic charge is necessary to have maximum conductivities, and existence of magnetic monopole with a critical charge (137 $e$) to reach the maximum superconductivity is important. Also, we show that thermal conductivity increases with increasing of magnetic charge. It may be concluded that the origin of superconductivity is magnetic monopole.}

\keywords{Superconductor; Black hole; Holography.}

\maketitle

\section{Introduction}
AdS/CFT correspondence, which relate a conformal field theory (CFT) in $d$-dimension and a string theory in $(d+1)$-dimensional anti-de Sitter (AdS) space, is a powerful mathematical tool to study strongly correlated systems \cite{P1, P2, P3}. The extra dimension in the AdS side can be understood as the energy scale of the CFT on the boundary. AdS/CFT correspondence already used to study quark-gluon plasma (QGP) \cite{P4, P5}, and it is possible to study nuclear force using AdS/CFT correspondence \cite{Nucl}. An interesting application of AdS/CFT is in condensed matter physics known as AdS/CMT which claim that there is a dual gravitational description of superconductivity \cite{P6, P7}, where properties of a (2+1)-dimensional
superconductor reproduced using a neutral black hole with a charged scalar and the only Maxwell sector. So, it can be generalize to the case with external magnetic field \cite{M01, M02, M03, M04}. In order to review some important aspects of the theory of holographic superconductors see Refs. \cite{P8, P9}.\\
Superconductivity is a phenomenon that has a history of about 100 years. Superconductivity is a state of a material in which $\rho = 0$ (disappearance of resistivity). It occurs only below a certain temperature ($T_{C}$), below a certain current level ($J_{C}$), and below a certain magnetic field ($H_{C}$).
Superconductivity was first discovered by Onnes in 1911 for mercury. Later, many researches has been done, theoretically and scientifically. The goal of the scientists was to justify this phenomenon by condensed matter and many-body models. From these kinds of models, we have Fritz and Heinz two fluid model (1935), phenomenological theories (QM) were proposed to explain  superconductivity by Ginzburg and Landau (1950), and cooper's BCS (Bardeen-Cooper-Schrieffer) pair model \cite{BCS}. But cooper's theory can't justify superconductivity in higher temperatures \cite{1, 2}.
On the other side, scientific works in this field are done in the basis of increasing the superconductivity transition temperature. In these days, scientists have reached a transition temperature of about 140K. But reaching higher temperatures is a great challenge for researchers.\\
In 1975, P. B. Price and E. K. Shirk \cite{Price} investigated about detecting of a moving magnetic monopole. They conducted an experiment, which with using a balloon-borne stack of Cherenkov film, emulsion, and 33 Lexan sheets, detected a very heavy particle passed through them. For this heavy particle as magnetic monopole, electrical charge of $Q=137e$ with mass of 200 proton mass was obtained.
This experiment is in agreement with investigates charged particles by V. R. Malkus \cite{MALKU} in 1951, for arbitrary magnetic moment, moving
simultaneously in the field of the monopole and an external electric field. It is concluded that the monopole can be coupled to matter with energies comparable to, but not significantly greater than, the chemical bond, reservations being made in the case of hydrogen where the lowest energy state depends upon the mass of the monopole. However, in 1982, the first results from a superconductive detector for moving magnetic monopoles, by Blas Cabrera, reported \cite{M}.
Considering a magnetic charge moving at velocity $v$ along the axis of a superconducting wire ring of certain radius, the velocity-and mass-independent search for moving magnetic monopoles is being performed by continuously monitoring the current in a 20-cm-area superconducting loop. A single candidate event, consistent with one Dirac unit of magnetic charge, has been detected during five runs.\\
The present work is done on the basis of a computing and analytical gravitation method and discusses the relation between magnetic monopole model and superconductivity. Heavy monopole initially suggested by 't Hooft and Polyakov in the framework of SU(2) gauge theories \cite{3, 4, 5}.
Our model tries to link the superconductivity to magnetic monopole charge carrier's mobility which have major effect on thermal and superconductive properties.\\
As we know, ordinary superconductors are well described by microscopic theory of superconductivity known as BCS theory \cite{BCS}. But, unusual superconductors including the pairing mechanism, is not good understood using BCS theory and one need another theoretical model of a strongly coupled system like AdS/CFT correspondence. In order to have realistic application of holography to
superconductivity it is important to shown how to introduce a dynamical (electromagnetic)
gauge field in holographic superconductors \cite{new1, new2, new3, new4}. In that case unusual magnetic materials with the strongly correlated effects has been studied with the holographic duality \cite{cai2}, while antiferromagnetism \cite{cai3} as well as paramagnetism \cite{cai4} quantum phase transition can be considered by using holographic principles. According to the AdS/CFT dictionary, a black hole and charged scalar field are holographic dual of temperature and condensate of a superconductor respectively \cite{P6}. In order to reproduce the superconductor phase diagram, one can consider a black hole with scalar hair at low temperature. STU black hole is an important model with both electric and magnetic charges \cite{1507.05553}. The STU model is just some $N=2$ supergravity \cite{STU1, STU2}. It generally involves 8 charges (4 electric and 4 magnetic). There is complete STU symmetry, so the "4" have "3" charges that are on the same footing, while the last has different couplings. The STU model can be interpreted in string theory by embedding. There are different ways to do this but the simplest, in type IIA on $T^2 \times T^2 \times T^2$, interprets the 8 charges as D0, D2 (3 of these, one for each $T^2$), D4 (3 of these as well) and D6. Special case of STU model in 5 dimensions with three electric charges which admits a chemical potential for the $U(1)^{3}$ symmetry already used to study QGP which is called STU/QCD correspondence \cite{B1, B2, B3, B4, B5}. Also $D=5$, $N=2$ STU model considered as dual picture of superfluid \cite{B6}.\\
Motivated by the evidence of superconductive detector for moving magnetic monopoles \cite{M}, we would like to investigate effect of magnetic charge on the conductivities. Special case of STU model in five dimensions already considered to study transport properties of superconductor \cite{0912.2228}. In that case, various kinds of STU model are important from AdS/CFT point of view and statistical analysis \cite{pourdarvish}, because it is extension of Yang-Mills theory to the case of having chemical potential. Recently, a deformation of the  $N = 2$, $D = 4$ STU model, characterized by a non-homogeneous special K\"{a}hler manifold, considered to solve BPS attractor equations and to construct static supersymmetric black holes with radial symmetry \cite{1507.05553}. In that case the relevant physical properties of the resulting black hole solution are explained and one can see that it is four charged STU model, three electric and one magnetic charges. So, we consider $N = 2$, $D = 4$ STU model as dual picture of a superconductor and investigate effect of magnetic charge on the transport properties. In another word, we wold like to claim that black hole magnetic charge is corresponding to the magnetic monopole in gauge theories.\\
This paper is organized as follows. In the next section, we review holographic superconductor and recall basic equations. Then, in section 3 we consider non-homogeneous STU black hole in four dimensions and write important properties of the model. In section 4 we discuss briefly about the thermodynamics of the model and calculate some useful quantities to study conductivities. In section 5 we study electrical end thermal conductivities and discuss about the effect of magnetic charge on them. Finally, in section 6 we give conclusion and summary of results together outlook of future works.

\section{Holographic superconductor}
It has been shown that a gravitational background may considered as holographically dual picture of a superconductor \cite{P6}, and it is known as holographic superconductor \cite{Musso}. It means that properties of strongly coupled superconductors in three dimensions may be described by four-dimensional classical general relativity.\\
Perturbation equation help us to obtain transport coefficients from the general Lagrangian of the form,
\begin{equation}\label{H1}
L\equiv\frac{\mathcal{L}}{\sqrt{-g}}=R-\frac{1}{4}G_{ij}F_{\mu\nu}^{i}F^{\mu\nu j}+\cdots,
\end{equation}
where $R$ stands for Ricci scalar and dotes denote scalar field and Chern-Simons terms, also $F_{\mu\nu}$ is the field-strength tensor. Induced metric $G_{ij}$ related to the background metric $g_{ij}$ with determinant $g$, which will have introduced later. In the Ref. \cite{0912.2228}, electrical and thermal conductivities of R-charged black hole in 4, 5, and 7 dimensions calculated for the general D-dimensional space-time,
\begin{equation}\label{H2}
ds^{2}=g_{tt} dt^{2}+g_{rr} dr^{2}+ g_{xx}d\Omega_{D-2}^{2},
\end{equation}
where $d\Omega^{2}$ is (D-2)-dimensional space. We will use results of the Ref. \cite{0912.2228} to study transport properties of superconductor via a non-homogeneous deformation of R-charged black hole (STU black hole) in four dimensions. Perturbation equation for the gauge field of general case can be written as follow \cite{0912.2228},
\begin{equation}\label{H3}
\frac{d}{dr}\left(N_{i}\frac{d}{dr}\phi_{i}(r)\right)-\omega^{2}N_{i}g_{rr}g^{tt}\phi_{i}(r)+\sum_{j=1}^{m}M_{ij}\phi_{j}(r)=0,
\end{equation}
where
\begin{equation}\label{H4}
N_{i}=\sqrt{-g}G_{ii}g^{xx}g^{rr},
\end{equation}
and
\begin{equation}\label{H5}
M_{ij}=F_{rt}^{i}\sqrt{-g}G_{ii}g^{xx}g^{rr}G_{jj}F_{rt}^{j},
\end{equation}
with the condition $M_{ij}=M_{ji}$. We can see that coupling of scalar field to the gauge field exists in the last term of the equation (\ref{H3}). Also, the scalar field $\phi_{i}$ related to the $x$-component of the gauge field via,
\begin{equation}\label{H55}
A_{x}^{i}=\frac{\mu^{i}}{2}\Phi_{i}(r)e^{i(qz-\omega t)},
\end{equation}
where
\begin{equation}\label{H55-1}
\phi_{i}(r)=\mu^{i}\Phi_{i}(r),
\end{equation}
with $\omega$ and $q$ represent the frequency and momentum in $z$ direction, respectively, and $\mu^{i}$ denotes the chemical potential. Hence, chemical potential (Maxwell charge) coupled to the scalar field $\Phi_{i}(r)$ at all.
It should be noted that the last term of the equation (\ref{H3}) comes from the metric perturbation part which interpreted as interaction between different gauge fields, so it will be vanish for the case of non-interacting fields. Also the second term of the equation (\ref{H3}) vanishes for the case of low frequency limit ($\omega^{2}\rightarrow0$).\\
The fact is that the Maxwell equations imply that the phase of the complex scalar field must be constant. Hence, without loss of generality one can consider scalar field as real variable.\\
The diagonal and off diagonal components of conductivity given by the following expressions respectively,
\begin{equation}\label{H6}
\sigma_{ii}=\frac{1}{8\pi G_{D}}\left[\sqrt{\frac{g_{rr}}{-g_{tt}}}\frac{\sum_{k=1}^{m}N_{k}(r)\phi_{k}(r,\omega)\phi_{k}(r,-\omega)}{(\phi_{i})^{0}(\phi_{i})^{0}}\right]_{r=r_{+}},
\end{equation}
and
\begin{equation}\label{H7}
\sigma_{ij}=\frac{1}{16\pi G_{D}}\left[\sqrt{\frac{g_{rr}}{-g_{tt}}}\frac{\sum_{k=1}^{m}N_{k}(r)\phi_{k}(r,\omega)\phi_{k}(r,-\omega)}{(\phi_{i})^{0}(\phi_{j})^{0}}\right]_{r=r_{+}},
\end{equation}
where $r_{+}$ is the horizon radius, and $G_{D}$ is $D$-dimensional Newtonian constant. We will use above relations to study electrical conductivity of a superconductor dual of deformed STU model in four dimensions. It should be noted that the general form of conductivities in absence of chemical potential obtained by the Ref. \cite{0809.3808}, but there is a Chern-Simons term in STU model and we have both electric and magnetic charges, and therefore chemical potential exists. So, one can use general expression which given by the Ref. \cite{0912.2228}.\\
On the other hand, coefficient of thermal conductivity given by,
\begin{equation}\label{H8}
\kappa_{T}=\left(\frac{\epsilon+P}{T}\right)^{2}\frac{1}{\sum_{i,j=1}^{m}\rho_{i}(\frac{iT}{\omega}G_{ij}(\omega))^{-1}\rho_{j}},
\end{equation}
where $\epsilon$ and $P$ are the local energy density and pressure respectively. Also, $\rho_{i}$ are charge density satisfy the following thermodynamics equations,
\begin{equation}\label{H9}
\epsilon+P=Ts+\sum_{i=1}^{m}\mu^{i}\rho_{i},
\end{equation}
and
\begin{equation}\label{H10}
d\epsilon=Tds+\sum_{i=1}^{m}\mu^{i}d\rho_{i},
\end{equation}
where $s$ and $\mu^{i}$ are entropy and chemical potential respectively.\\
It should be noted here that there are two order parameters, the first is the coherence length ($\zeta$), which is the scale of order parameter variation. It is indeed corresponding to the radial coordinate $r$ and hence $\phi(r)$ will be complex order parameter to realize a holographic
p-wave model \cite{cai5, cai6}. The second is the London penetration depth ($\lambda$), which characterizing the magnetic field penetration to superconductor. It may be corresponding to magnetic charge $p^{0}$ which will be introduce in the next section.

\section{Non-homogeneous STU black hole in four dimensions}
As we know, STU background is extension of AdS-Schwarzschild background (which is holographic dual of ordinary Yang-Mills theory) to include chemical potential due to presence of the black hole charge \cite{B1}. Hence, we expect that the 4D deformed background of STU model could be holographic dual of a superconductor as well as ordinary AdS-Schwarzschild black hole background \cite{P6}. The main reason which we choose this special background is presence of magnetic charge. We would like to study effect of the magnetic charge on superconductivity. As the background considered here is solution of $N=2$, $D=4$ gauged supergravity coupled to Abelian vector multiples, it can reflect some properties of strongly coupled superconductors in 2+1-dimensional space-time.\\
Non-homogeneous deformation of the STU model of $N=2$, $D=4$ gauged supergravity has been studied by the Ref. \cite{1507.05553}, and we review some important properties of the model which will be useful to study holographic superconductor.
The general form of the metric given by,
\begin{equation}\label{S1}
ds^{2}=-U(r) dt^{2}+\frac{dr^{2}+\psi(r)d\Omega_{2}^{2}}{U(r)},
\end{equation}
where $d\Omega_{2}^{2}$ is the two-dimensional metric of surface. Determinant of above line element is,
\begin{equation}\label{S2}
\sqrt{-g}=\frac{\psi(r)}{U(r)}f_{k},
\end{equation}
with $k=0, \pm1$, where $f_{1}=\sin\theta$, $f_{0}=\theta$ and $f_{-1}=\sinh\theta$, which are corresponding to closed, flat and open universes.\\
The bosonic Lagrangian given by \cite{1507.05553},
\begin{equation}\label{S3}
L=R-g_{ij}\partial_{\mu}z^{i}\partial^{\mu}\bar{z}^{j}-\frac{1}{4}\mathcal{I}_{\Lambda\Sigma}F_{\mu\nu}^{\Lambda}F^{\mu\nu \Sigma}+\cdots,
\end{equation}
where
\begin{eqnarray}\label{S4}
\mathcal{I}_{\Lambda\Sigma}=\frac{1}{4}e^{-\mathcal{K}}\left(\begin{array}{ccc}
1 & 0\\
0 & 4g_{ij}\\
\end{array}\right).
\end{eqnarray}
with
\begin{eqnarray}\label{S5}
g_{ij}=\frac{1}{2(\lambda^{1}\lambda^{2}\lambda^{3}-\frac{A}{3}(\lambda^{3})^{3})^2} \left(\begin{array}{ccc}
(\lambda^{2})^{2}(\lambda^{3})^{2} & \frac{A}{3}(\lambda^{3})^{4}& -\frac{2A}{3}\lambda^{2}(\lambda^{3})^{3}\\
\frac{A}{3}(\lambda^{3})^{4} & (\lambda^{1})^{2}(\lambda^{3})^{2}& -\frac{2A}{3}\lambda^{1}(\lambda^{3})^{3}\\
-\frac{2A}{3}\lambda^{2}(\lambda^{3})^{3} & -\frac{2A}{3}\lambda^{1}(\lambda^{3})^{3}& (\lambda^{1})^{2}(\lambda^{2})^{2}+\frac{A^{2}}{3}(\lambda^{3})^{4}\\
\end{array}\right),
\end{eqnarray}
and
\begin{equation}\label{S6}
e^{-\mathcal{K}}=8\left(\lambda^{1}\lambda^{2}\lambda^{3}-\frac{A}{3}(\lambda^{3})^{3}\right),
\end{equation}
where $A$ is deformation parameter which may be fix as unit, so $A=0$ give us ordinary STU model. The superpotential specified by the dyonic Fayet-Iliopoulos (FI) gauging with FI parameters $(g^{\Lambda},g_{\Lambda})$. Also, in this model both magnetic ($p^{\Lambda}$) and electric ($q^{\Lambda}$) charges are possible. We will consider only the following non-zero charges, $p^{0}$, $q_{1}$, $q_{2}$ and $q_{3}$ which means that we have one magnetic and three electric charges. We are interested to the case of positive magnetic charge and consider it as our physical case, while for the electrical charge we are free to choose any positive or negative values. This choice requires ($g_{0}p^{0}-g^{i}q_{i}=-k$, where $k=0, \pm1$) that some FI parameters vanish, so we have only $g^{1}$, $g^{2}$, $g^{3}$ and $g_{0}$ which interpreted as gauge coupling constant (all positive) and related to the scalar fields $\lambda^{1}$, $\lambda^{2}$ and $\lambda^{3}$ which will mention later. In that case, full black hole solution of the non-homogeneous STU model given by,
\begin{equation}\label{S7}
\psi(R)=(ar^{2}-c)^{2},
\end{equation}
and
\begin{equation}\label{S8}
U(r)=\frac{2g_{0}g^{3}(ar^{2}-c)^{2}}{\lambda_{\infty}^{3}(ar-g_{0}\beta^{0}-\frac{g_{0}\beta_{3}}{(\lambda_{\infty}^{3})^{2}})
\sqrt{(ar+2g_{0}\beta^{0})(ar+\frac{2g_{0}\beta_{3}}{(\lambda_{\infty}^{3})^{2}})}},
\end{equation}
where $a$ (with the length dimension) and $c$ (with the inverse of length dimension) are positive constant and
\begin{equation}\label{S9}
\lambda_{\infty}^{3}=\sqrt{\frac{g_{0}g^{3}}{g^{1}g^{2}-\frac{A}{3}(g^{3})^{2}}},
\end{equation}
is asymptotic value of $\lambda^{3}$. Also, $\beta^{0}$ and $\beta_{i}$ with $i=1,2,3,$ are constants \cite{0911.4926}. It is found that,
\begin{eqnarray}\label{S10}
\lambda^{1}&=&\frac{a\frac{g^{1}}{g^{3}}(\lambda_{\infty}^{3})^{2}r-g_{0}\beta_{3}(\frac{g^{1}}{g^{3}}-A\frac{g^{3}}{g^{2}})-\beta^{0}\frac{(g_{0})^{2}}{g^{2}}}
{\sqrt{(2g_{0}\beta^{0}+ar)(2g_{0}\beta_{3}+ar(\lambda_{\infty}^{3})^{2})}},\nonumber\\
\lambda^{2}&=&\frac{g^{2}}{g^{3}}\lambda^{3},\nonumber\\
\lambda^{3}&=&\lambda_{\infty}^{3}\sqrt{\frac{ar+\frac{2g_{0}\beta_{3}}{(\lambda_{\infty}^{3})^{2}}}{ar+2g_{0}\beta^{0}}}.
\end{eqnarray}
In that case all quantities described by $\beta^{0}$, $\beta_{i}$, $a$, $c$, $p^{0}$ and $q_{i}$.\\
Black hole horizon given by
\begin{equation}\label{S11}
r_{+}=\sqrt{\frac{c}{a}}.
\end{equation}
Now, we can discuss about the perturbation equation (\ref{H3}) and obtain $\phi_{i}$ with $i=0,1,2,3$ corresponding to one magnetic and three electric charges of black hole.\\

\begin{figure}[h!]
\begin{center}$
\begin{array}{cccc}
\includegraphics[width=55 mm]{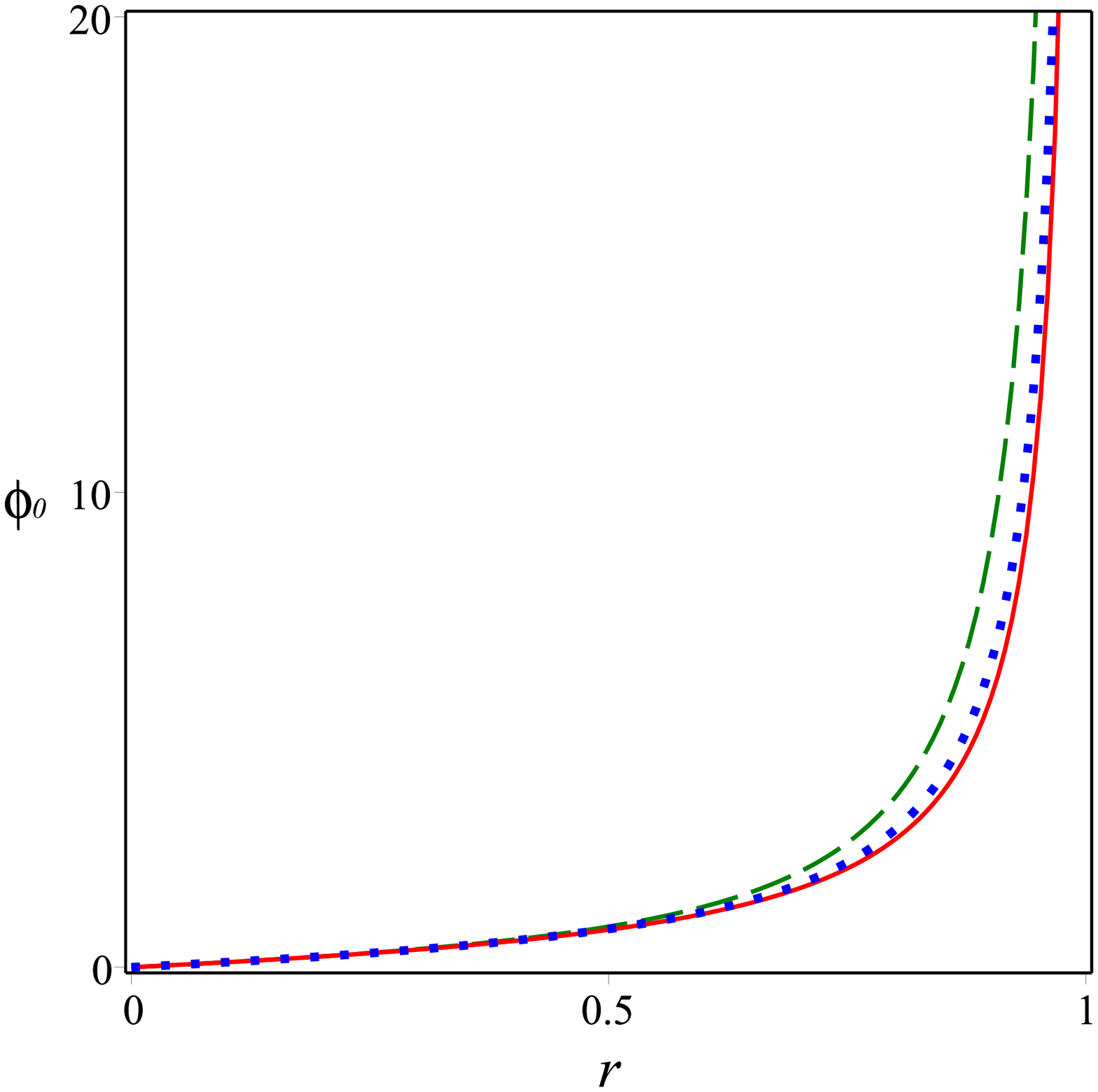}&\includegraphics[width=55 mm]{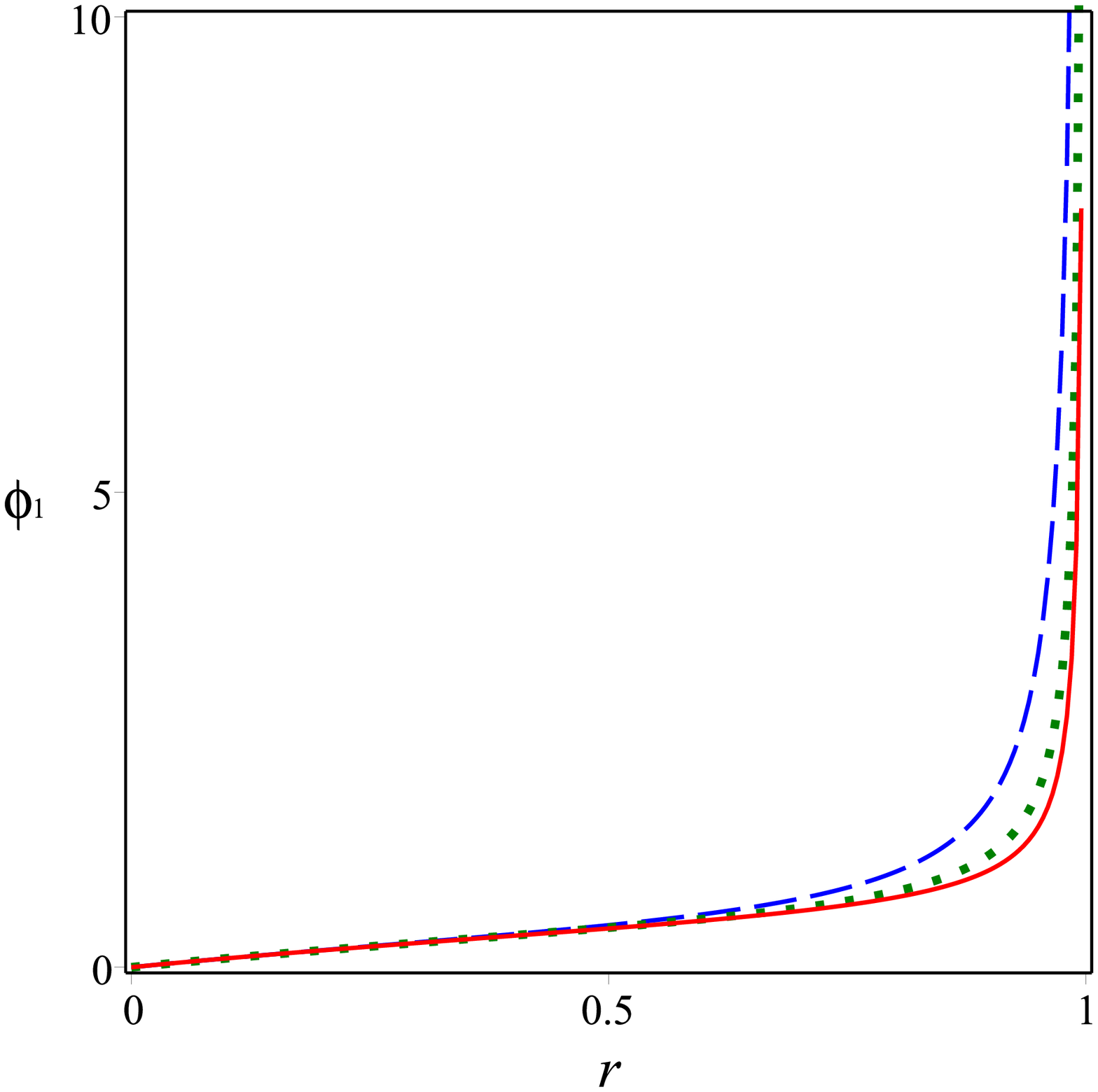}\\
\includegraphics[width=55 mm]{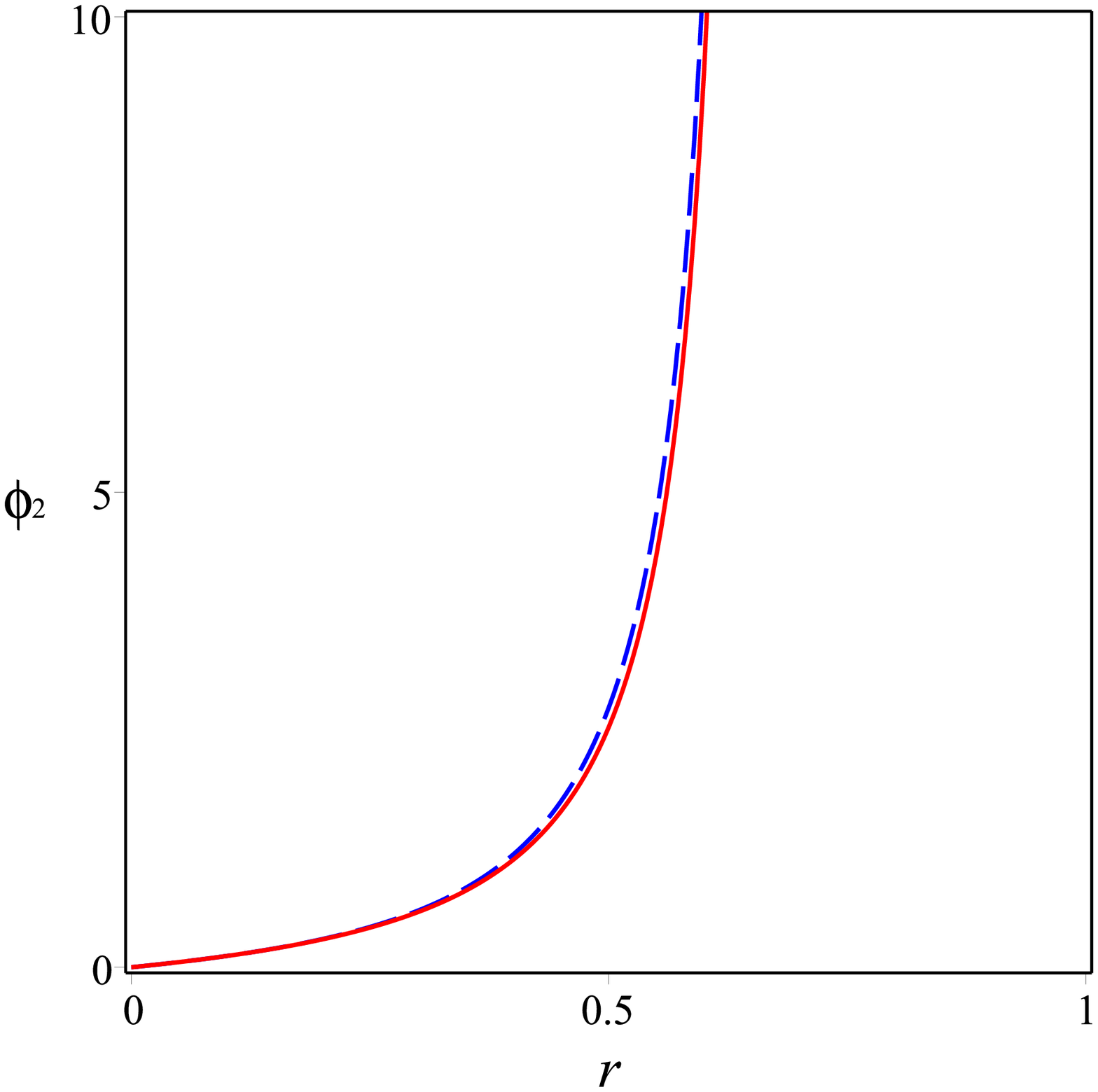}&\includegraphics[width=55 mm]{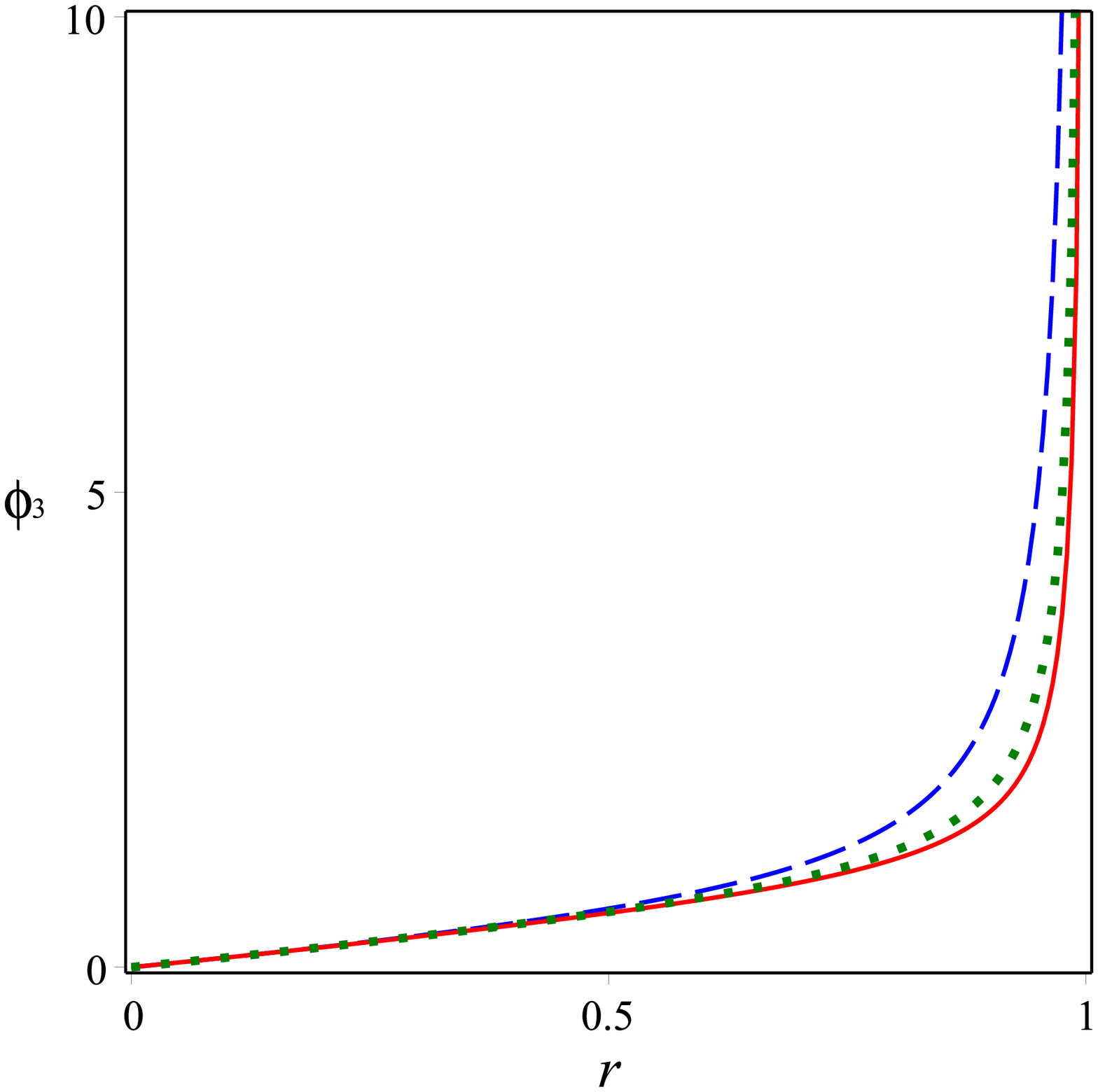}
\end{array}$
\end{center}
\caption{Typical behavior of scalar fields in terms of $r$ ($r\leq r_{+}$) with initial values $\phi(0)=0$ and $\phi^{\prime}(0)=1$. All model parameters set one ($a=c=A=g_{0}=g^{1}=g^{2}=g^{3}=\beta^{0}=\beta_{3}=f_{k}=1$). $\omega=1$ (dashed blue), $\omega=0.5$ (dotted green), $\omega=1$ (solid red).}
\label{fig:1}
\end{figure}

\begin{figure}[h!]
\begin{center}$
\begin{array}{cccc}
\includegraphics[width=55 mm]{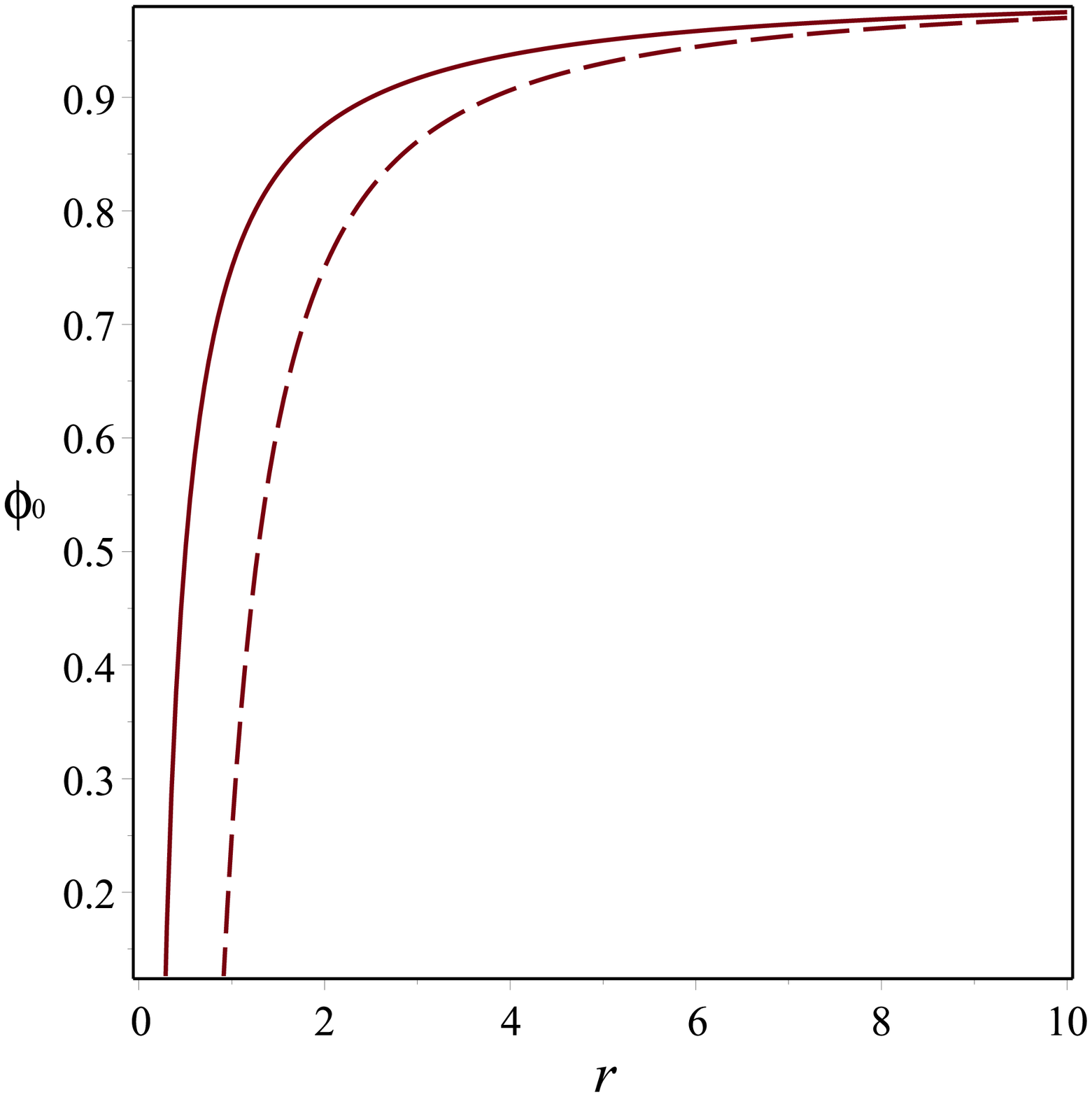}&\includegraphics[width=55 mm]{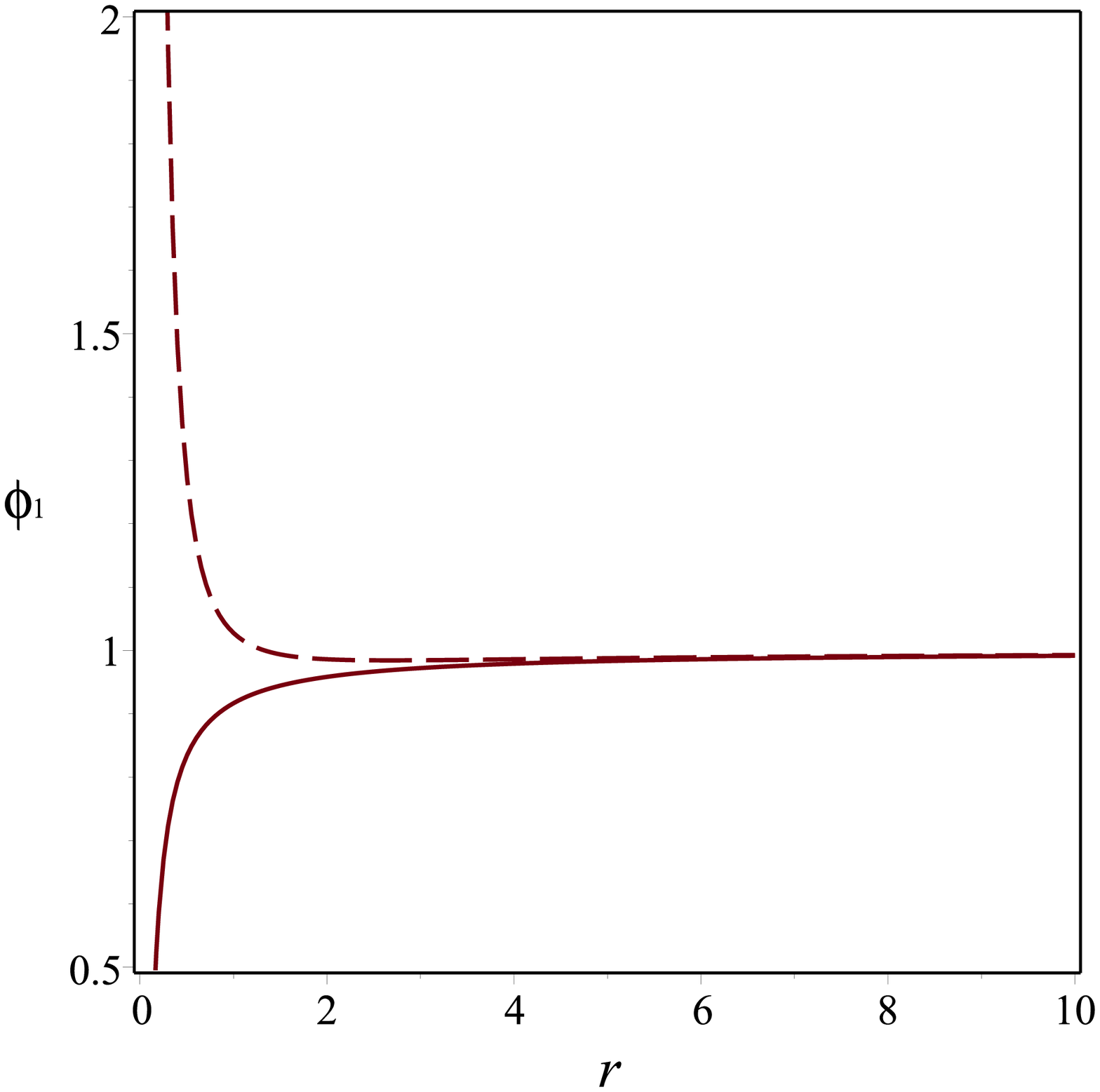}\\
\includegraphics[width=55 mm]{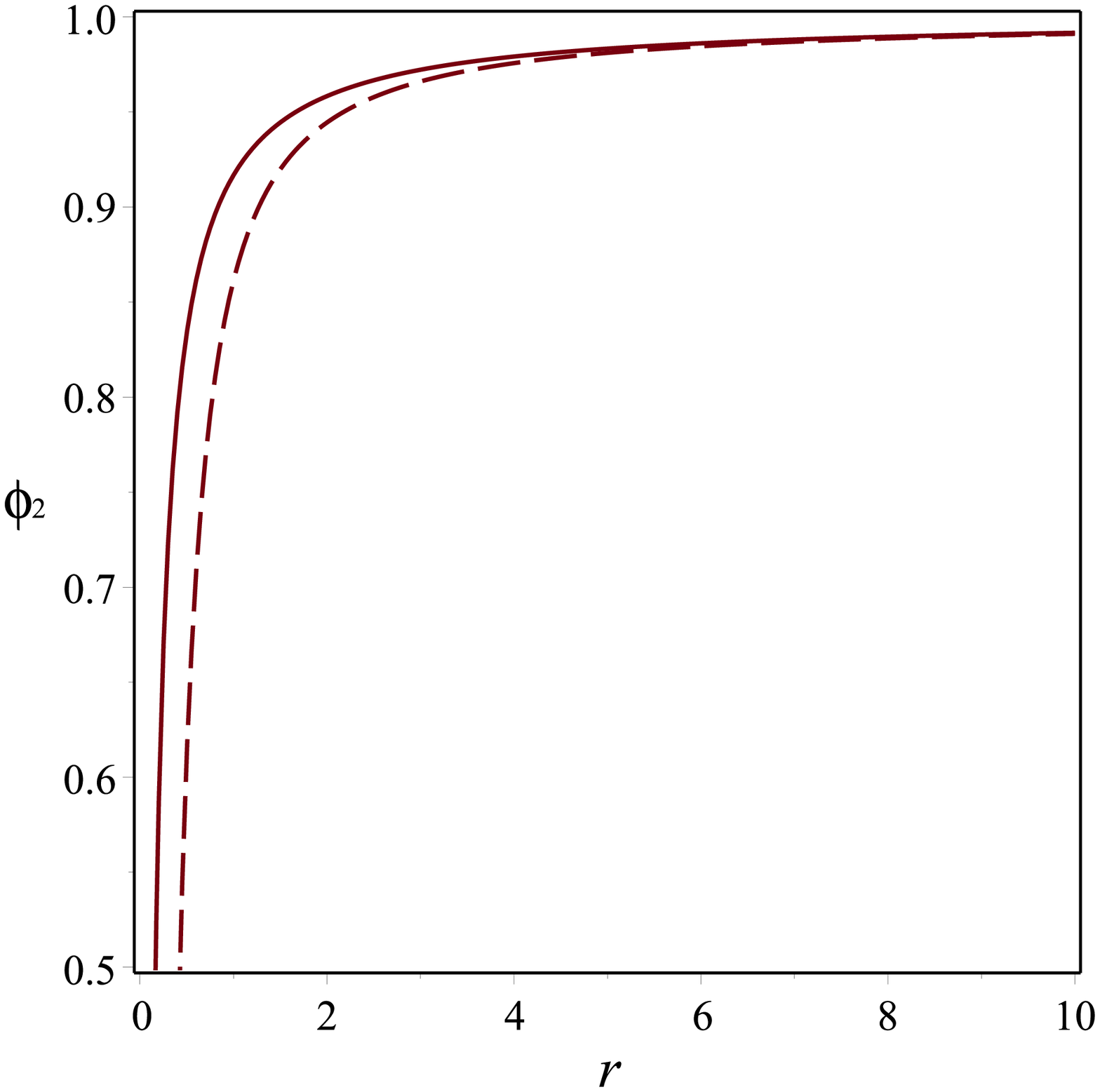}&\includegraphics[width=55 mm]{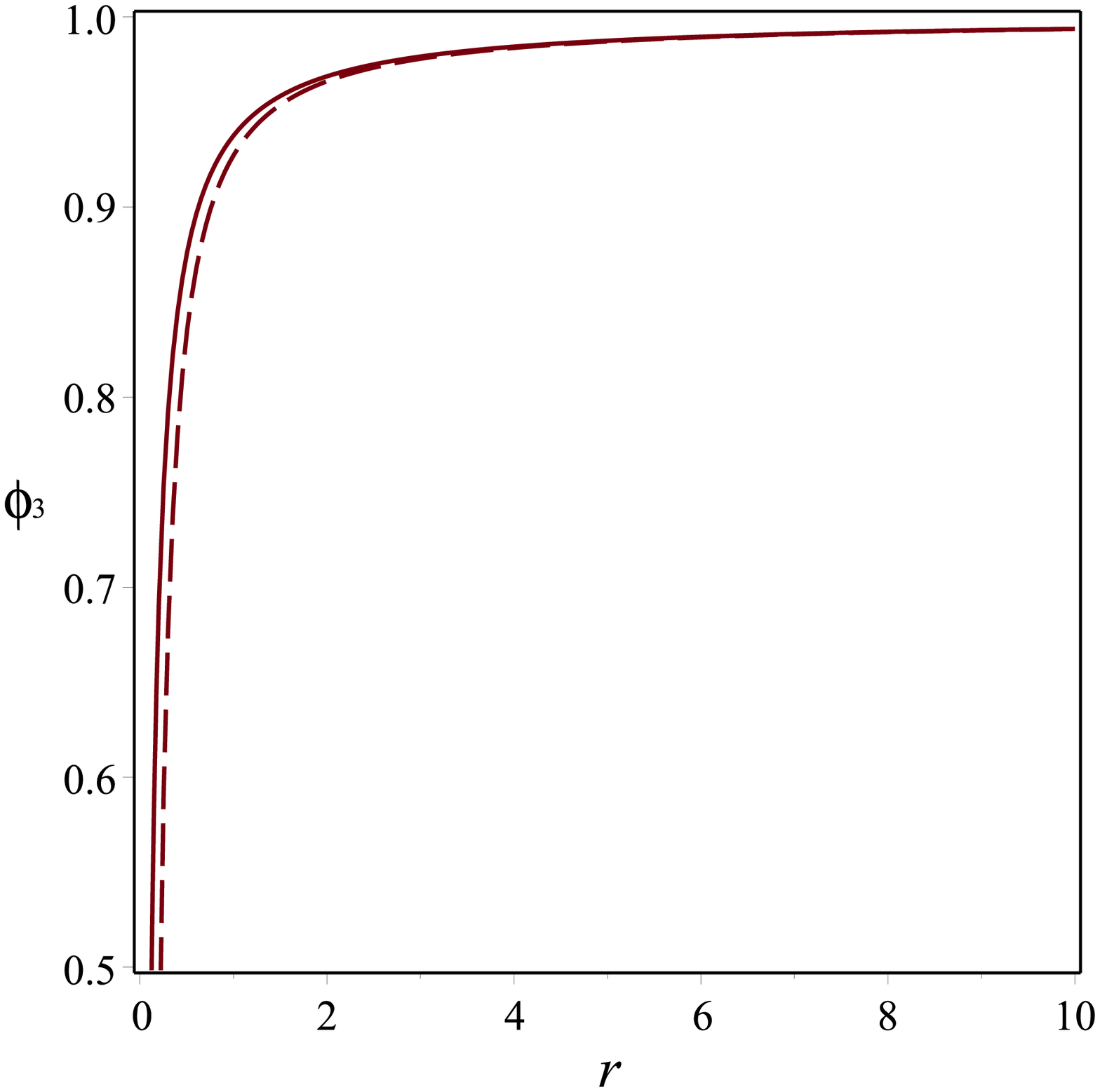}
\end{array}$
\end{center}
\caption{Typical behavior of scalar fields in terms of $r$ ($r\geq r_{+}$) for $\omega^{2}=0$. All model parameters set one ($a=c=A=g_{0}=g^{1}=g^{2}=g^{3}=\beta^{0}=\beta_{3}=f_{k}=C_{1}=C_{2}=1$). Solid lines denote first order approximation while dashed lines represent second order approximation.}
\label{fig:2}
\end{figure}

Using the equation (\ref{H4}) it is easy to find that,
\begin{equation}\label{S12}
N_{0}=2f_{k}\left(\lambda^{1}\lambda^{2}\lambda^{3}-\frac{A}{3}(\lambda^{3})^{3}\right)U(r),
\end{equation}
\begin{equation}\label{S12-2}
N_{1}=\frac{4f_{k}(\lambda^{2})^{2}(\lambda^{3})^{2}}{\left(\lambda^{1}\lambda^{2}\lambda^{3}-\frac{A}{3}(\lambda^{3})^{3}\right)}U(r),
\end{equation}
\begin{equation}\label{S12-3}
N_{2}=\frac{4f_{k}(\lambda^{1})^{2}(\lambda^{3})^{2}}{\left(\lambda^{1}\lambda^{2}\lambda^{3}-\frac{A}{3}(\lambda^{3})^{3}\right)}U(r),
\end{equation}
and
\begin{equation}\label{S12-4}
N_{3}=\frac{4f_{k}\left((\lambda^{1})^{2}(\lambda^{2})^{2}+\frac{A^{2}}{3}(\lambda^{3})^{4}\right)}{\left(\lambda^{1}\lambda^{2}\lambda^{3}-\frac{A}{3}
(\lambda^{3})^{3}\right)}U(r),
\end{equation}
where $U(r)$ given by the equation (\ref{S8}).
Then, non-interacting version of the equation (\ref{H3}) reads as,
\begin{equation}\label{S13}
\frac{d}{dr}\left(N_{i}\frac{d}{dr}\phi_{i}(r)\right)+\omega^{2}N_{i}\frac{\phi_{i}(r)}{U(r)}=0.
\end{equation}
So, we have four different equation to obtain $\phi_{0}(r)$, $\phi_{1}(r)$, $\phi_{2}(r)$, and $\phi_{3}(r)$. In the simplest case of low frequency we can neglect $\omega^{2}$. In that case one can obtain,
\begin{equation}\label{S14}
\phi_{i}(r)=C_{1}\int{N_{i}^{-1}dr}+C_{2},
\end{equation}
where $C_{1}$ and $C_{2}$ are integration constants.\\
Numerically, we solve equation (\ref{S13}) and obtain behavior of $\phi_{i}$ with respect to $r$ ($r<r_{+}$) as illustrated by the plots of Fig. \ref{fig:1}. We find that presence of all FI parameters are necessary to obtain finite and real value of scalar fields. It is illustrated that scalar fields asymptotically yields to infinity near the black hole horizon ($r_{+}=1$ for $a=c=1$). We have shown that $\phi_{2}>\phi_{0}>\phi_{3}>\phi_{1}$. It is clear that $\phi_{i}$ are increasing function of $r$ inside black hole. Behavior of scalar fields are in agreement with the results of the Ref. \cite{0912.2228} in the region of $r<r_{+}$. We also find that effect of deformation if decreasing value of scalar field. It means that for the bigger $A$ we have smaller $\phi_{i}$.\\
It is desirable that scalar field should be well-defined outside horizon with positive value. At the low frequency limit, one can use the equation (\ref{S14}) and obtain asymptotic values of the scalar fields for the large $r$ to find behavior of scalar field as illustrated by the Fig. \ref{fig:2}. We obtain plots corresponding to $A=1$ but situation is similar for $A=-1$ which used in the Ref. \cite{0912.2228}. We will show later that, negative value is suitable for the deformation parameter $A$ as used in the Ref. \cite{0912.2228}. So the boundary value of the scalar field $(\phi_{i})^{0}$ is a constant (may be zero or one), for example $(\phi_{i})^{0}\approx1$ for the given value of model parameters as Fig. \ref{fig:1} and Fig. \ref{fig:2}.

In agreement with the Ref. \cite{0912.2228} and coincide with both large and small $r$ we can propose the following function for the scalar field,
\begin{equation}\label{S14-1}
\phi_{i}(r)=\frac{a_{i}+b_{i}r}{c_{i}+d_{i}r},
\end{equation}
with $a_{i}<0$, $b_{i}<0$, $c_{i}<0$ and $d_{i}>0$. It is indeed corresponding to $C_{2}=0$ of the solution (\ref{S14}). However we can consider it as general behavior of the scalar field satisfying the differential equation (\ref{S13}) for non-interacting case with $\omega^{2}\rightarrow0$.
\section{Thermodynamics}
In this section we will study thermodynamics properties of our system near the equilibrium. Near the horizon we have infinitesimal temperature which is decreasing function of $r$, and the black hole entropy given by \cite{0912.2228},
\begin{equation}\label{S15}
s=\frac{\pi}{3}\frac{\mathcal{A}}{p^{0}},
\end{equation}
where the black hole horizon area given by,
\begin{equation}\label{S16}
\mathcal{A}=2\left(-p^{0}q_{3}\left[(-p^{0}q_{3})^{2}+12(p^{0})^{2}q_{1}q_{2}\right]+\left[(-p^{0}q_{3})^{2}-4(p^{0})^{2}q_{1}q_{2}\right]^{\frac{3}{2}}\right),
\end{equation}
with,
\begin{equation}\label{S17}
p^{0}=\frac{ac}{2g_{0}}-2g_{0}(\beta^{0})^{2},
\end{equation}
as magnetic charge and,
\begin{eqnarray}\label{S18}
q_{1}&=&2(\beta_{3})^{2}\frac{g^{2}}{(g^{3})^{2}}\left(g^{1}g^{2}-\frac{A}{3}(g^{3})^{2}\right)-g^{2}\frac{ac}{2\left(g^{1}g^{2}-\frac{A}{3}(g^{3})^{2}\right)},\nonumber\\
q_{2}&=&\frac{1}{2g^{2}}(\beta^{0}g_{0}+\beta_{3}\frac{g^{1}g^{2}}{g^{3}})^{2}-g^{1}\frac{ac}{2(g^{1}g^{2}-\frac{A}{3}(g^{3})}\nonumber\\
&+&\frac{A}{3}\beta_{3}\frac{g^{3}}{g^{2}}\left(\beta_{3}\frac{g^{1}g^{2}}{g^{3}}-\beta^{0}g_{0}-\frac{A}{2}\beta_{3}g^{3}\right),\nonumber\\
q_{3}&=&\frac{g^{2}}{g^{3}}q_{2}-A\frac{g^{3}}{g^{2}}q_{1},
\end{eqnarray}
as electric charges.\\
In the case of positive magnetic charge ($p^{0}\geq0$), the equation (\ref{S17}) requires that $g_{0}\leq\frac{ac}{4(\beta^{0})^{2}}$. If we choose unit value for the parameters (as selected in figures), then we should have $g_{0}\leq0.5$. So we use this as allowed regions in the figures.\\

\begin{figure}[h!]
\begin{center}$
\begin{array}{cccc}
\includegraphics[width=55 mm]{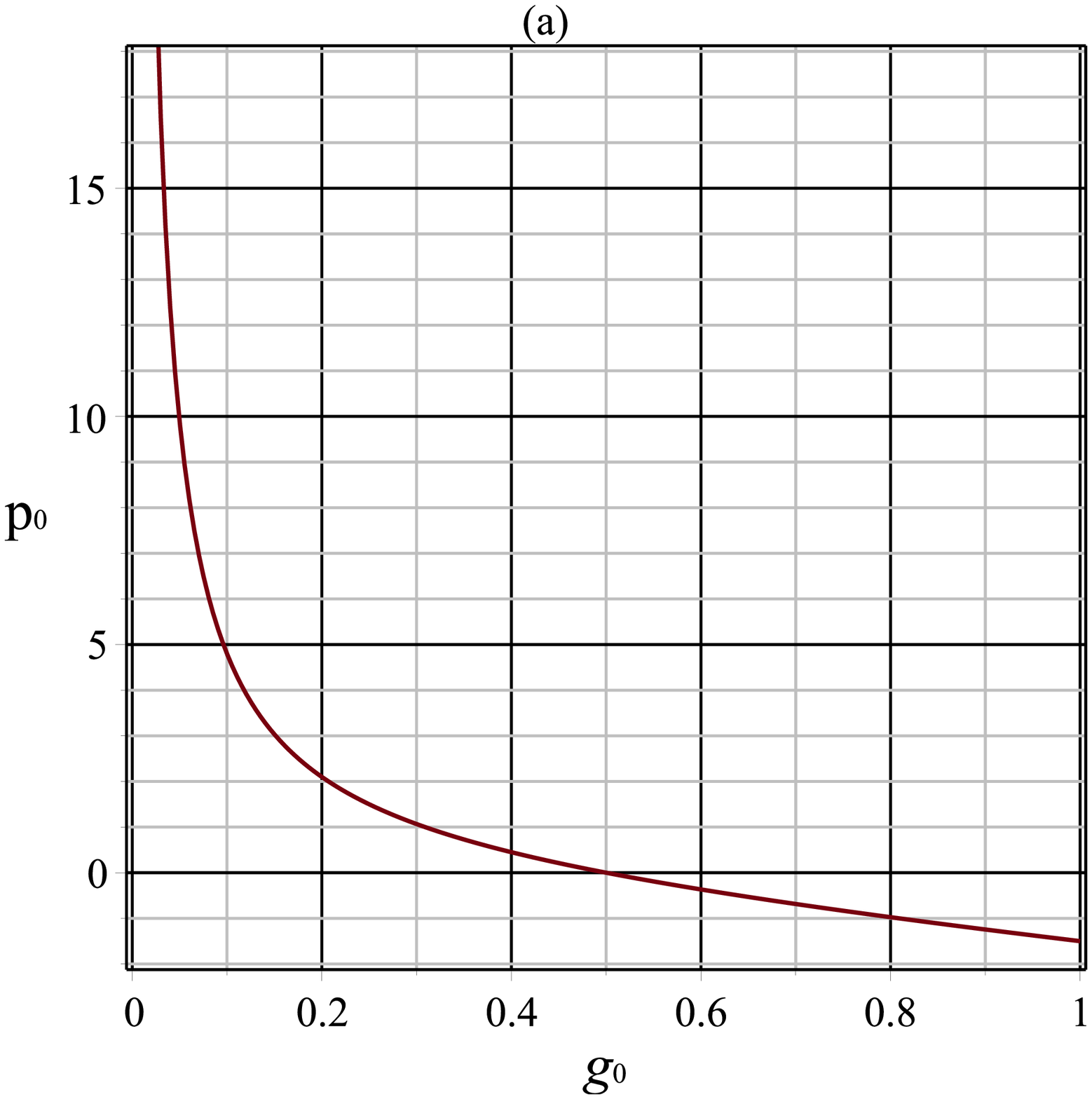}&\includegraphics[width=55 mm]{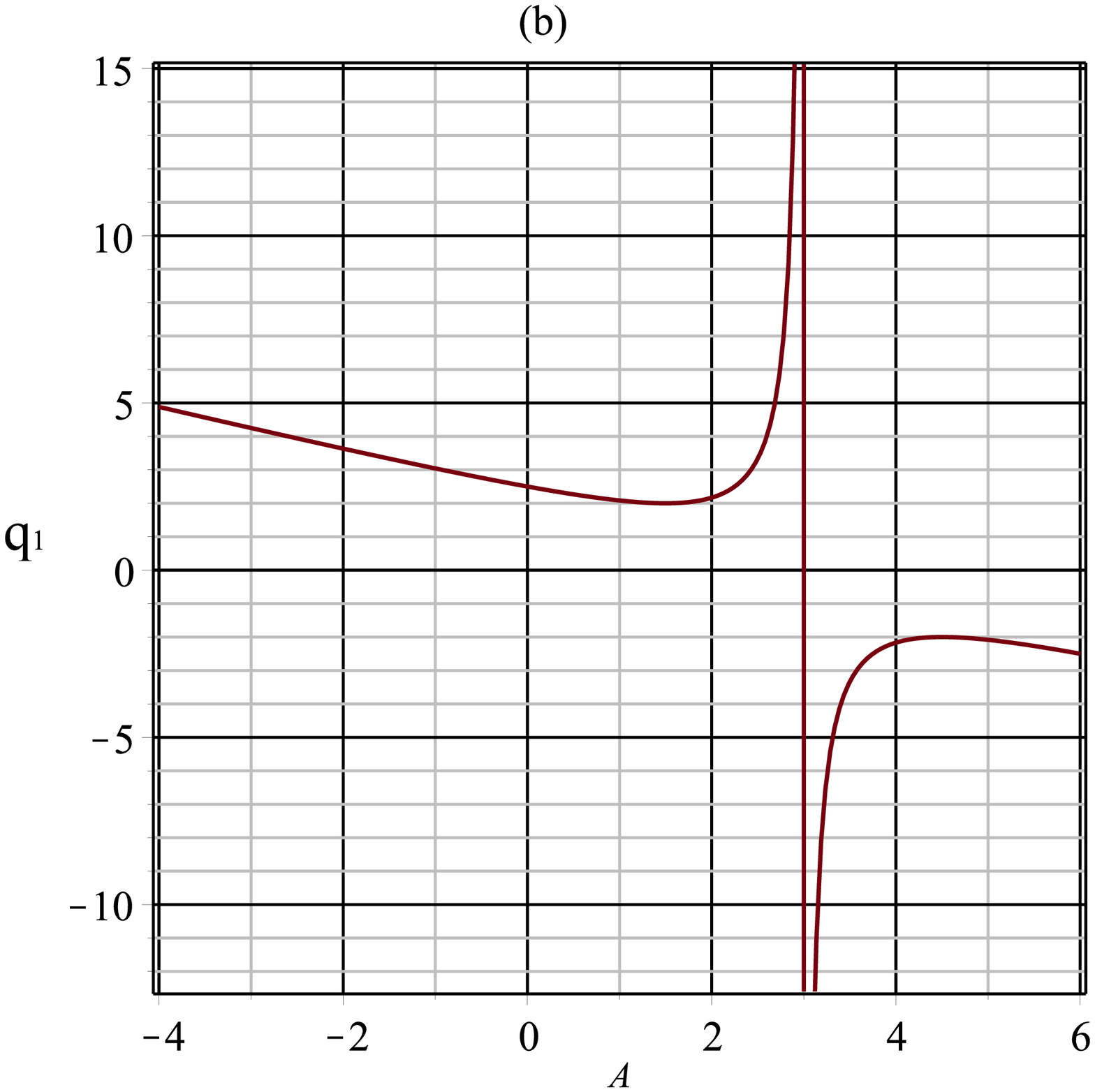}\\
\includegraphics[width=55 mm]{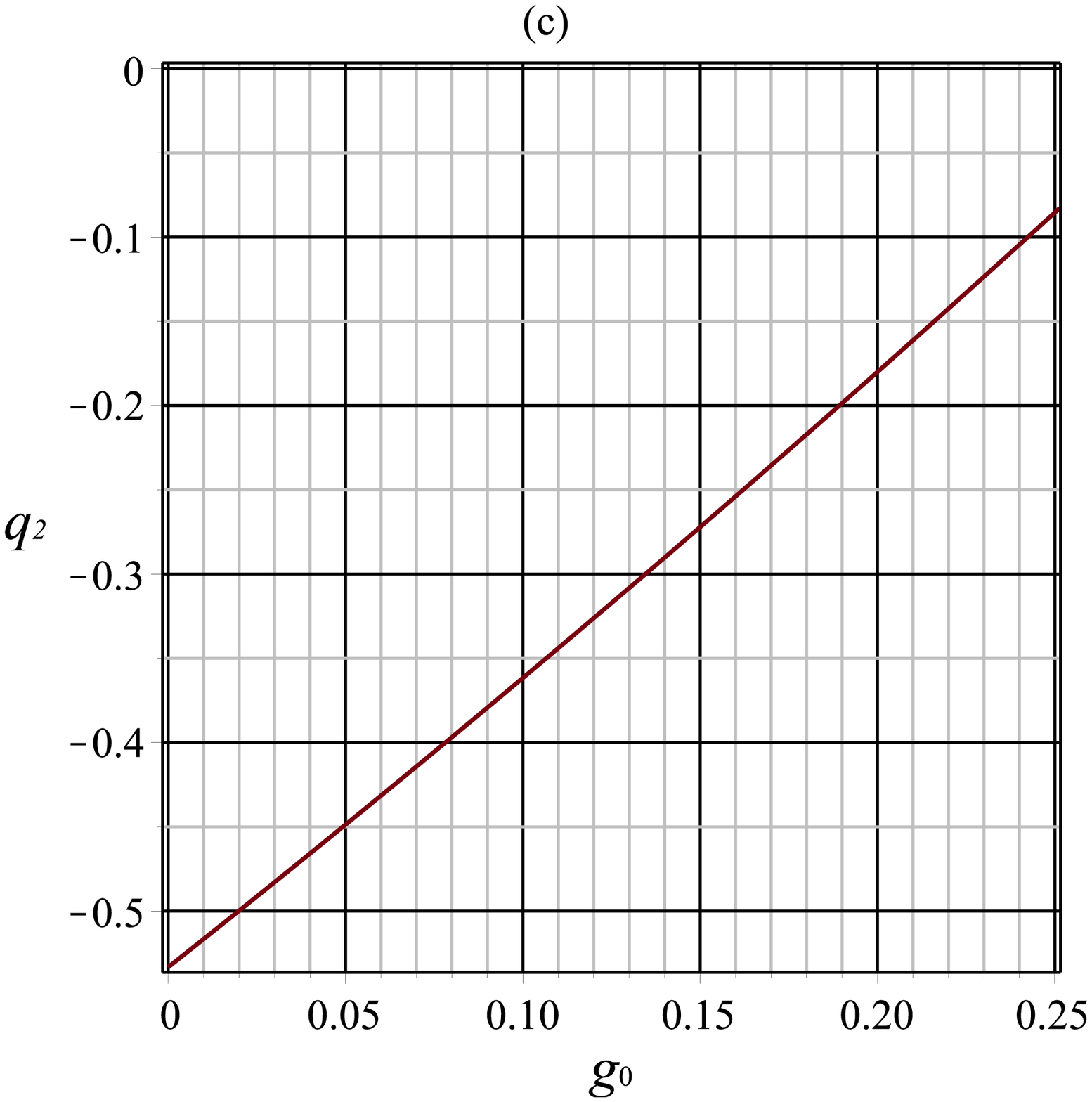}&\includegraphics[width=55 mm]{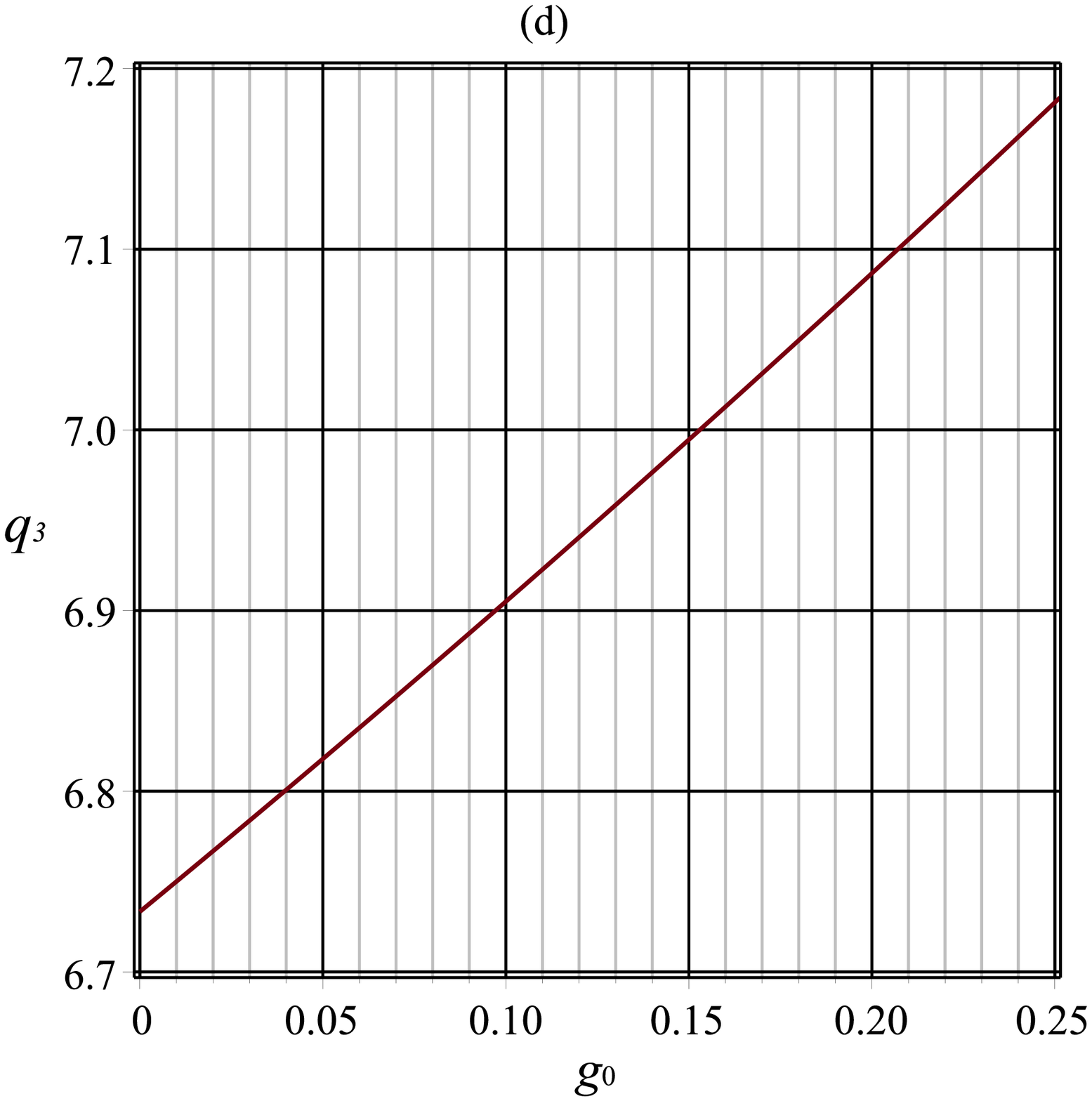}
\end{array}$
\end{center}
\caption{Magnetic and electric charges in terms of $g_{0}$ for $a=c=g^{1}=g^{2}=g^{3}=\beta^{0}=\beta_{3}=1$ and $A=-2$.}
\label{charges}
\end{figure}
In the Fig. \ref{charges} we show values of magnetic and electric charges for the case of $a=c=\beta^{0}$ corresponding to $r_{+}=1$. According to the Fig. \ref{charges} (a) it is clear that infinitesimal $g_{0}$ yields to large positive value for the magnetic charge while large value of $g_{0}$ yields to large negative value for the magnetic charge, which is not our physical case (see equation (\ref{S17})) and we only focus on the regions between $0\leq g_{0}\leq0.5$. It is clear that magnetic charge vanishes for $g_{0}=\pm\frac{\sqrt{ac}}{2\beta^{0}}$. As initial assumption FI parameters all are positive, so positive sign is acceptable and $g_{0}=\frac{\sqrt{ac}}{2\beta^{0}}$ is corresponding to zero magnetic charge. In the case of $a=c=\beta^{0}$ we have $g_{0}=0.5$ where $p^{0}=0$. In the case of $g_{0}=0.1$ we have $p^{0}=5$.\\
In the Fig. \ref{charges} (b) we draw $q_{1}$ as a function of deformation parameter $A$ because $q_{1}$ is not depend on $g_{0}$. Minimum value of $q_{1}$ obtained for $A\approx1.5$ (positive charge) and $A\approx4.5$ (negative charge). In the case of $A=-1$ we have $q_{1}=3$ and $A=-2$ we have $q_{1}\approx3.5$. For the negative $A$ we have positive $q_{1}$ as well as $0\leq A<3$, while for the $A>3$ we have negative $q_{1}$.\\
Again in the Fig. \ref{charges} (c) we draw $q_{2}$ in terms of $g_{0}$ and see that $g_{0}<0.3$ yields to negative electric charge while for the case of $g_{0}>0.3$ we have positive $q_{2}$ which is increasing function of $g_{0}$. Hence, in the allowed region of $g_{0}$ we may have both positive or negative $q_{2}$.\\
In the Fig. \ref{charges} (d) we can see that $q_{3}$ is totally positive for positive $g_{0}$ and it is increasing function of $g_{0}$. In the case of $g_{0}=0.1$ we have $q_{3}=6.9$.

\begin{figure}[h!]
\begin{center}$
\begin{array}{cccc}
\includegraphics[width=75 mm]{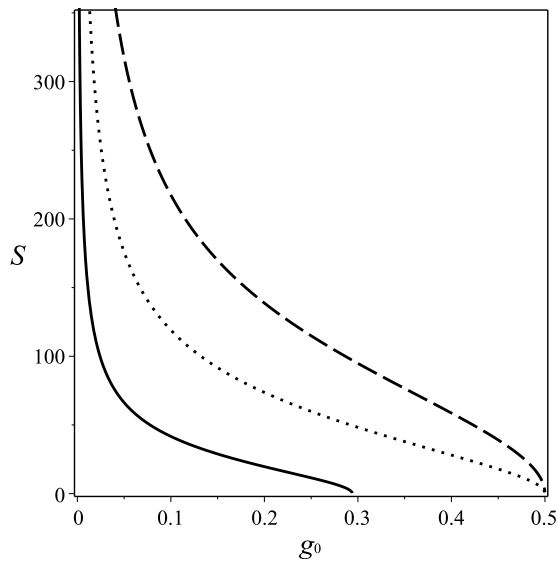}
\end{array}$
\end{center}
\caption{Typical behavior of entropy in terms of $g_{0}$ for $a=c=g^{1}=g^{2}=g^{3}=\beta^{0}=\beta_{3}=1$. $A=-2$ Solid, $A=-3$ dotted, $A=-4$ dashed.}
\label{fig:3}
\end{figure}

\begin{figure}[h!]
\begin{center}$
\begin{array}{cccc}
\includegraphics[width=75 mm]{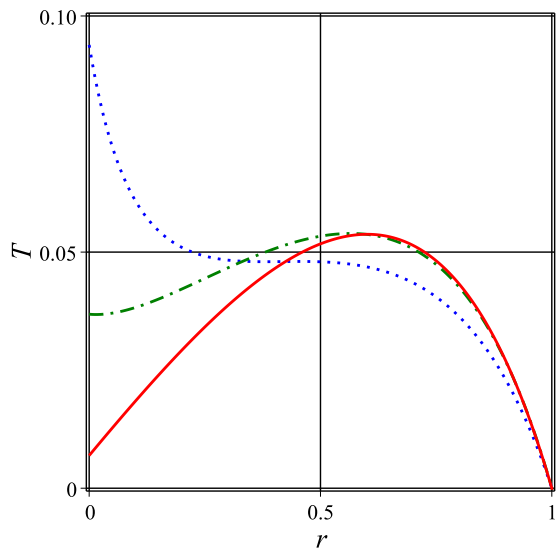}
\end{array}$
\end{center}
\caption{Typical behavior of local temperature in terms of $r$ for $a=c=g^{1}=g^{2}=g^{3}=\beta^{0}=\beta_{3}=1, A=-2$. $g_{0}=0.1$ blue dot, $g_{0}=0.2$ green dash dot, $g_{0}=0.5$ (zero magnetic charge) red solid.}
\label{fig:4}
\end{figure}

In the Fig. \ref{fig:3} we draw entropy (\ref{S15}) and see that it is decreasing function of $g_{0}$ and yields to constant value for large $g_{0}$. For the selected value of model parameter, we should choose negative $A$ to have positive entropy. Analyzing the specific heat tell that the black hole is in the stable phase. We investigate thermodynamics quantities near the horizon, and we need local temperature of the black hole which is drawn in terms of $r$ by the Fig. \ref{fig:4}.
We can see that black hole temperature is zero at $r=r_{+}=1$ as expected and it is decreasing function near the horizon. For the small value of $g_{0}$ (large positive value of magnetic charge), the black hole temperature is totally decreasing function of $r$. For the larger value of $g_{0}$, the black hole temperature increases first to a maximum value, then decreases to zero at the black hole horizon.
Zero magnetic charge case represented by solid red line of the Fig. \ref{fig:4}.
In that case, the local temperature is approximately like parabola, hence we can write local temperature as a function of $r$.
\section{Conductivities}
In this section we try to obtain electrical and thermal conductivities. Obtaining transport properties depending on the black hole charges (specially magnetic charge) are the main goal of this paper. In order to obtain electrical conductivity we assume low frequency limit ($\omega^{2}\rightarrow0$) of the metric (\ref{S1}) and use the scalar field (\ref{S14-1}) in the relations (\ref{H6}) and (\ref{H7}).
\subsection{Electrical conductivity}
\begin{figure}[h!]
\begin{center}$
\begin{array}{cccc}
\includegraphics[width=60 mm]{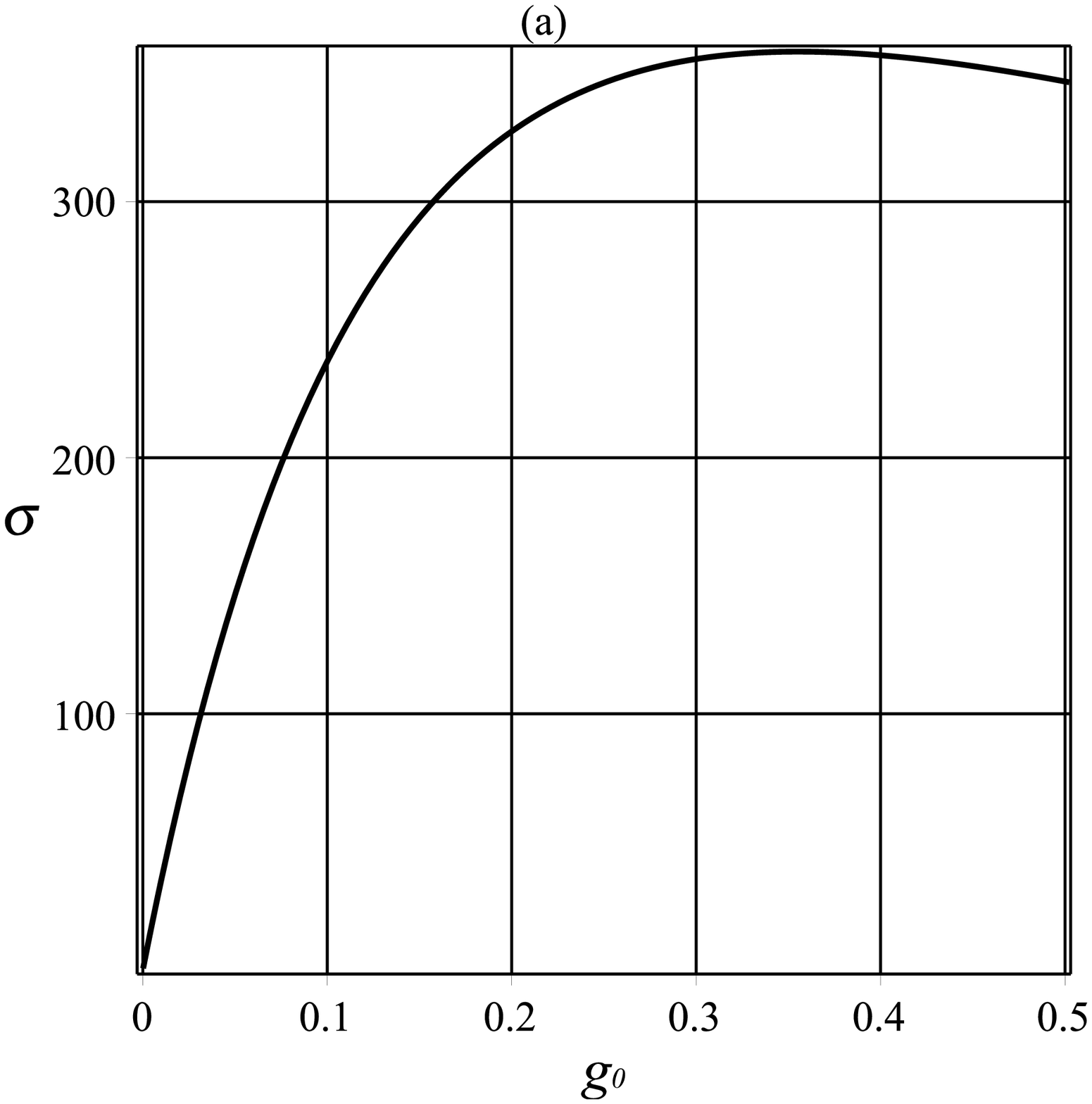}&\includegraphics[width=60 mm]{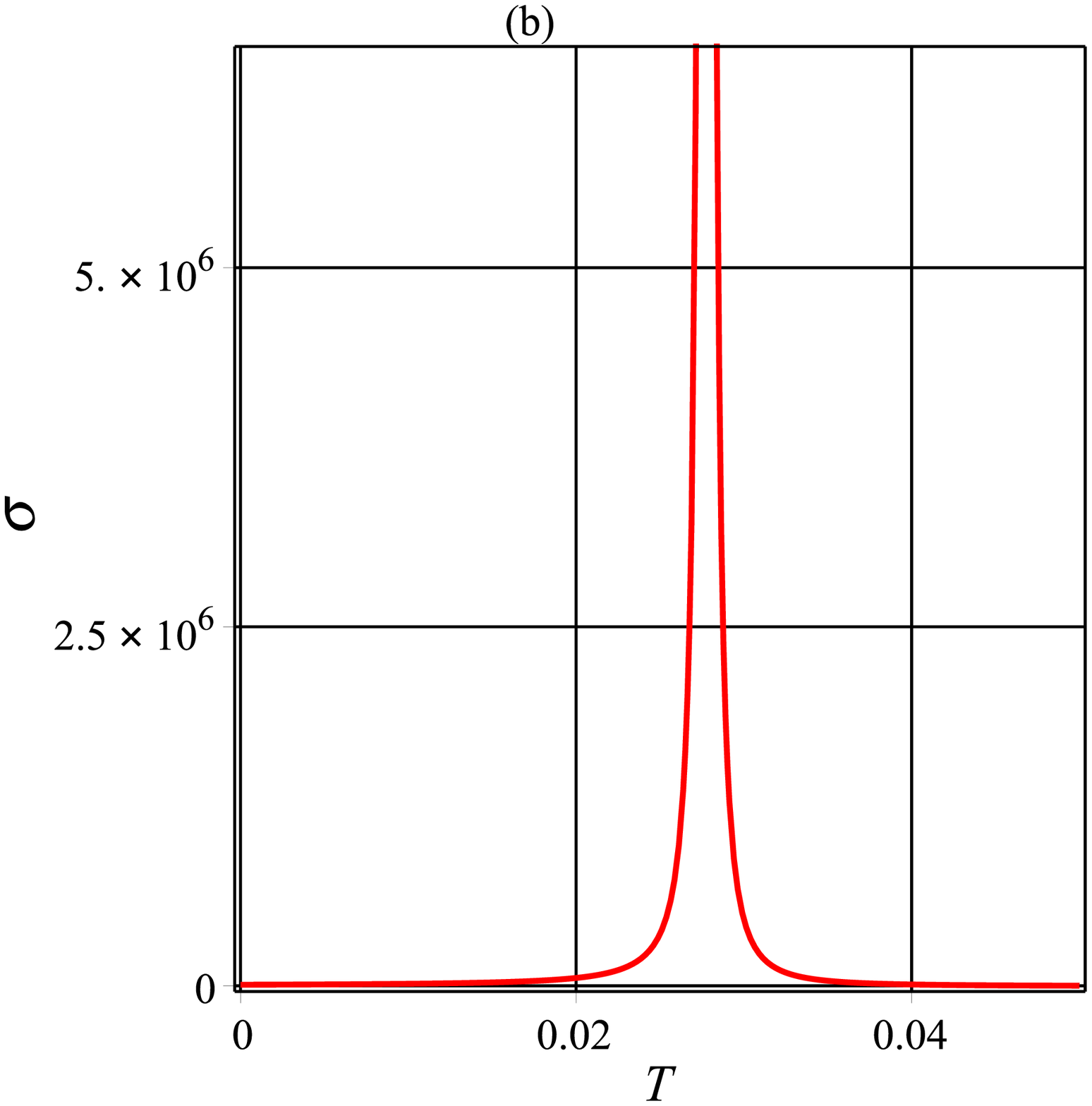}
\end{array}$
\end{center}
\caption{Typical behavior of electrical conductivity for $\omega^{2}=0, a=c=g^{1}=g^{2}=g^{3}=\beta^{0}=\beta_{3}=f_{k}=1$, and $A=-2$. (a) in terms of $g_{0}$ with $A=-2$ and $r=1$. (b) in terms of $T$ with $A=1.4$ and $g_{0}=0.3$.}
\label{fig:6}
\end{figure}

Assuming $(\phi_{i})^{0}=(\phi_{j})^{0}$ give us similar diagonal and off diagonal conductivity components, hence $\sigma=\sigma_{ii}=\sigma_{ij }$. In the unit of $8\pi G=1$ and $m=1$ we can obtain near horizon behavior of electrical conductivity components, graphically, as illustrated by plots of the Fig. \ref{fig:6}. At $T_{C}$, there is a second order phase transition from a normal metal into a superconducting state that is much like critical phenomena such as superfluities, magnetic ultra- thin films  and superconductors.\\
It is clear that electrical conductivity increased below phase transition temperature while decreased at higher temperature. It means that we have high conductivity at low temperature as expected.\\
Also, in the Fig. \ref{fig:6} (a) we can find the effect of magnetic charge on the electrical conductivity near the horizon. In the case of $A=-2$, it has been shown that maximum conductivity given by $g_{0}\approx0.35$ which means $p^{0}\approx0.5$, $q_{1}\approx3.5$, $q_{2}\approx0.1$ and $q_{3}\approx 7.4$  (see Fig. \ref{charges}).\\
It means that, in order to have maximum conductivity, presence of magnetic charge is necessary. However, decreasing or increasing magnetic charge decreases value of electrical conductivity. It means that there is a critical magnetic charge where the maximum of conductivity exists. By using results of the Fig. \ref{fig:4} we can fit temperature as a function of $r$ and then obtain plot of conductivity in terms of temperature (see Fig. \ref{fig:6} (b)). Asymptotic behavior in the Fig. \ref{fig:6} (b) show superconductor phase transition in presence of magnetic charge.\\

\begin{figure}[h!]
\begin{center}$
\begin{array}{cccc}
\includegraphics[width=65 mm]{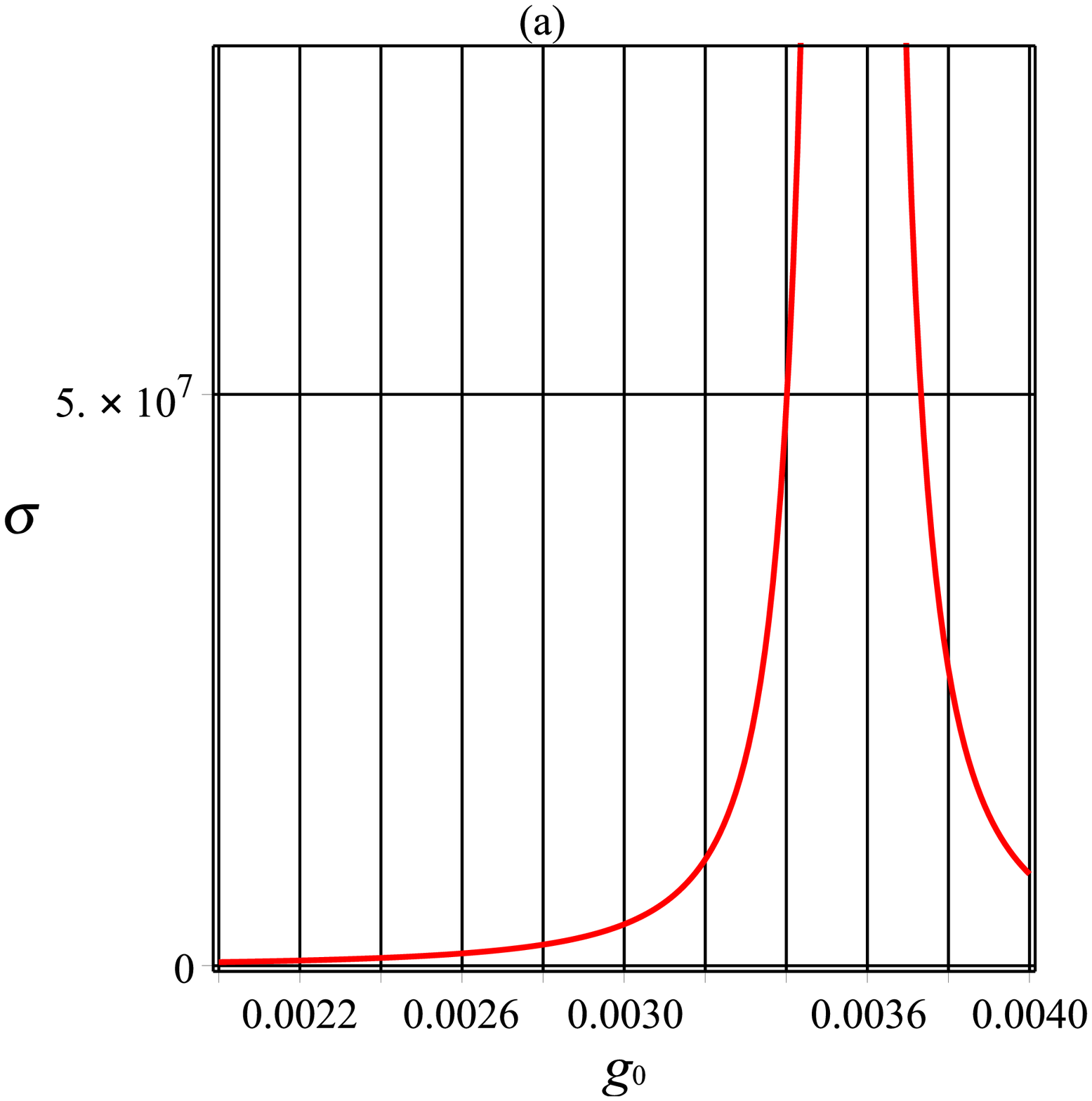}&\includegraphics[width=65 mm]{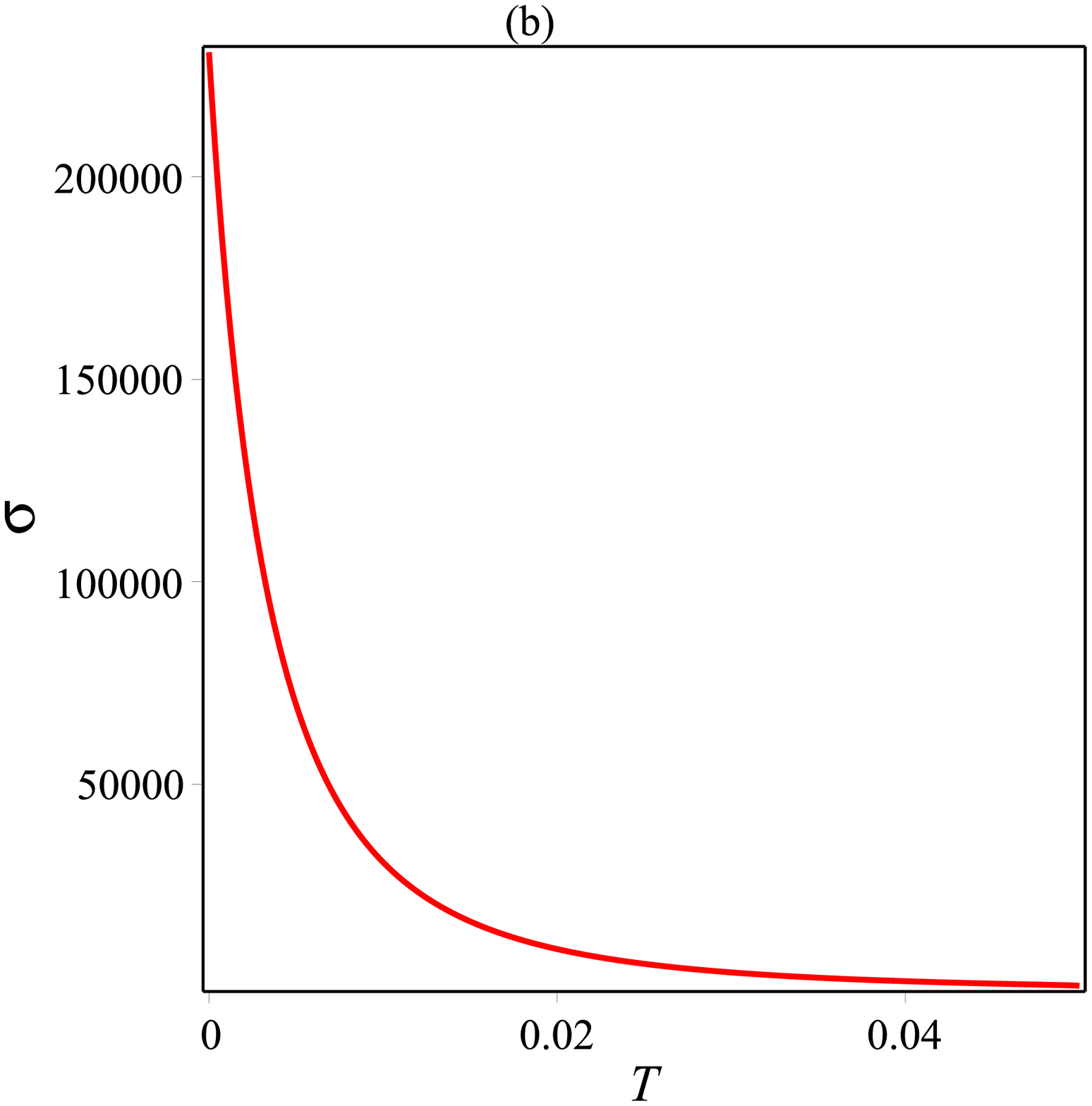}
\end{array}$
\end{center}
\caption{Typical behavior of electrical conductivity for $\omega^{2}=0, a=c=g^{1}=g^{2}=g^{3}=\beta^{0}=\beta_{3}=f_{k}=1$, (a) in terms of $g_{0}$ with $A=0$ and $r=1$. (b) in terms of $T$ with $A=1.4$ and $g_{0}=0.5$ (zero magnetic charge).}
\label{fig:7}
\end{figure}

However, deformation parameter is also important quantity in conductivity. We find, for small negative deformation parameter, that superconductivity happen for the large positive value of magnetic charge. It means that for the infinitesimal value of deformation parameter, superconductivity is due to large value of magnetic charge. In another word, superconductivity enhanced near horizon due to the magnetic charge for the case of ordinary STU model ($A=0$). In this case ($A=0$), we find special cases of superconductivity (see Fig. \ref{fig:7} (a)) with $g_{0}\approx0.0036$ where $q_{2}=q_{3}\approx1$ ($q_{1}=2.5$) and $p_{0}\approx137$. In that case $q_{2}$ and $q_{3}$ interpreted as electron charge so value of magnetic charge is about 137 electron charge. Comparison of Fig. \ref{fig:6} (b) and Fig. \ref{fig:7} (b), which are conductivities in terms of temperature, show that we can have superconductive phase transition by using appropriate choice of $A$ and $g_{0}$. These results is in agreement with other experimental results \cite{Bardeen}.\\
In the Fig. \ref{fig:7} (b) we can see superconductor at $T=0$ with zero magnetic charge, while in presence of magnetic charge we have superconductive phase transition at finite temperature. In summary, one can conclude that origin of superconductivity may be magnetic charge.

\subsection{Thermal conductivity}
\begin{figure}[h!]
\begin{center}$
\begin{array}{cccc}
\includegraphics[width=65 mm]{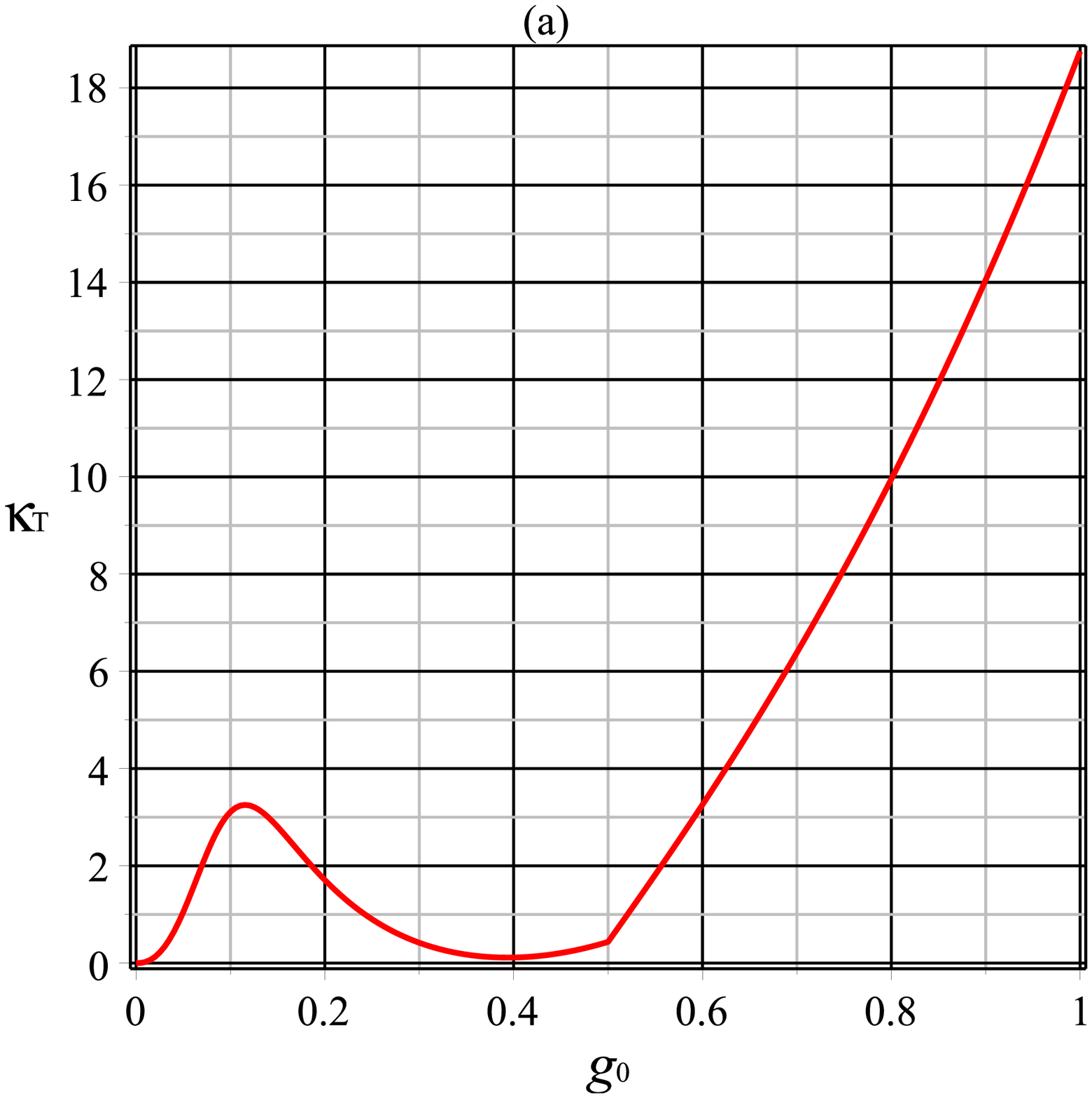}\includegraphics[width=65 mm]{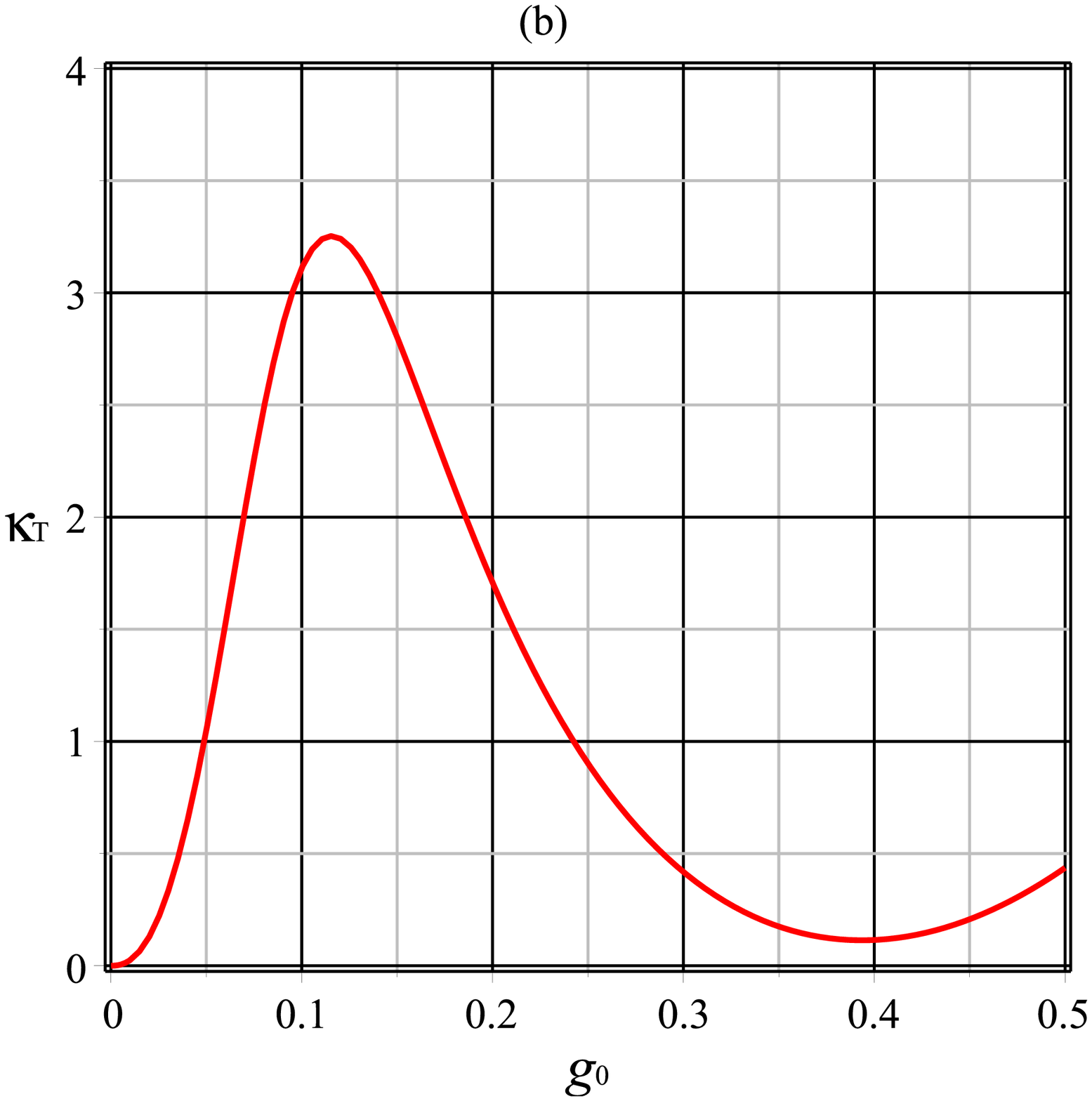}
\end{array}$
\end{center}
\caption{Thermal conductivity in terms of $g_{0}$ for $\omega^{2}=0, a=c=g^{1}=g^{2}=g^{3}=\beta^{0}=\beta_{3}=f_{k}=1, A=-2$, with near horizon behavior ($r\approx1$). (a) in the extended region, and (b) in the allowed region.}
\label{fig:8}
\end{figure}

In order to obtain thermal conductivity using the equation (\ref{H8}) we use the fact that chemical potential of STU black hole is proportional to the temperature $\mu^{i}\propto T$ \cite{1206.2067}, also we assume that $iT\approx\omega$, because $T\ll1$ and $\omega^{2}\ll1$. In that case we can rewrite thermal conductivity as follow,
\begin{equation}\label{H8-1}
\kappa_{T}=\frac{s^{2}+(\sum_{i=0}^{3}\rho_{i})^{2}}{\sum_{i,j=0}^{3}\rho_{i}(G_{ij}(\omega))^{-1}\rho_{j}},
\end{equation}
where the equation (\ref{H9}) is used. In the Fig. \ref{fig:8} (a) we draw thermal conductivity in terms of $g_{0}$ to see effect of magnetic charge. We find that deformation parameter $A$ should be negative to have real thermal conductivity. We find that the thermal conductivity is proportional to the temperature. It means that decreasing temperature reduces value of the thermal conductivity as expected. Effect of magnetic charge on the thermal conductivity is strange. In both cases of large positive and small positive magnetic charge there is no thermal conductivity. On the other hand, thermal conductivity increased dramatically for the case of large negative magnetic charge. However, it is not physical case, because we found from the equation (\ref{S17}) that $g_{0}\leq0.5$ (for the unit values of all parameters as selected in figures), hence in the Fig. \ref{fig:8} (b) we can see behavior of the thermal conductivity in the allowed region of $g_{0}$. We can see maximum of thermal conductivity for $g_{0}\approx0.12$ which means $p^{0}\approx4.8$. It means that finite value of magnetic monopole can yields to highest thermal conductivity.

\section{Conclusion}
In this paper, we considered non-homogeneous deformation of STU model in four dimensions including three electric and one magnetic charges and use fluid/gravity correspondence to study holographic superconductor. We obtained transport quantities like electrical and thermal conductivities and studied the effect of magnetic charge on them. We focus on the near horizon temperature which is corresponding to low temperature field theory. Numerically, we found behavior of the scalar field in terms of the radial coordinate which can be used to interpret temperature dependent behavior of the scalar field. In that case near horizon temperature is decreasing function of the radial coordinate. In order to investigate the effect of the electric and magnetic charges we fixed all parameters and varies only one related parameters which is $g_{0}$. This is one of the order parameter which is corresponding to London penetration depth, while coherence length is the other order parameter which is corresponding to the radially dependent scalar field. Our numerical analysis have shown that increasing of $g_{0}$ yields to decreasing of the magnetic charge and vice versa. Hence, we found that the entropy increases by increasing of the magnetic charge. We find that presence of the magnetic charge is necessary to have high electrical and thermal conductivities, and it may be related to  time-dependent perturbations which can enhance superconductivity \cite{N}. Small positive value of magnetic charge is enough to have maximum of electrical conductivity. Using the numerical analysis, we found critical value of the magnetic charge where there is maximum of the electrical conductivity. We have shown that ordinary STU model ($A=0$) is better model to describe holographic superconductor, however there are some situations with positive $A$ including superconductive phase transition.\\
In order to obtain thermal conductivity we used ordinary entropy given by (\ref{S15}), while it is interesting to calculate thermal conductivity with logarithmic corrected entropy \cite{log1,log2,log3,log4,log5,log6,log7} and study thermal fluctuations which is corresponding to $N^{-1}$ corrections in AdS/CFT.

\end{document}